# Room Temperature Magnetic Skyrmions in Gradient-Composition Engineered CoPt Single Layers

*Adam Erickson,[1][†] Qihan Zhang,[2][†] Hamed Vakili,[3][†] Chaozhong Li,[4] Suchit Sarin,[1] Suvechhya Lamichhane,[3] Lanxin Jia,[2] Ilja Fescenko,[5] Edward Schwartz,[3] Sy-Hwang Liou,[3] Jeffrey E. Shield,[1] Guozhi Chai,[4] Alexey A. Kovalev,[3] Jingsheng Chen,[2,6,*] Abdelghani Laraoui[1,3*]*

[1]Department of Mechanical & Materials Engineering, University of Nebraska-Lincoln, 900 N 16th Street, W342 NH, Lincoln, NE 68588, United States
[2]Department of Materials Science and Engineering, National University of Singapore, Block E2, #05-19, 5 Engineering Drive 2, Singapore 117579, Singapore
[3]Department of Physics and Astronomy and the Nebraska Center for Materials and Nanoscience, University of Nebraska-Lincoln, 855 N 16th St, Lincoln, NE 68588, United States
[4]Key Laboratory for Magnetism and Magnetic Materials of Ministry of Education, School of Physical Science and Technology, Lanzhou University, Lanzhou 730000, China
[5]Laser Center, University of Latvia, Jelgavas St 3, Riga, LV-1004, Latvia
[6]National University of Singapore (Suzhou) Research Institute, Suzhou, Jiangsu, 215123, China
[†]Equal contributions
[*]Corresponding authors: msecj@nus.edu.sg, alaraoui2@unl.edu

**ABSTRACT**

Topologically protected magnetic skyrmions in magnetic materials are stabilized by interfacial or bulk Dzyaloshinskii-Moriya interaction (DMI). Interfacial DMI decays with increase of the magnetic layer thickness in just a few nanometers and bulk DMI typically stabilizes magnetic skyrmions at low temperatures. Consequently, more flexibility in manipulation of DMI is required for utilizing nanoscale skyrmions in energy efficient memory and logic devices at room temperature (RT). Here, we demonstrate the observation of RT skyrmions stabilized by gradient DMI (g-DMI) in composition gradient engineered CoPt single layer films by employing topological Hall effect, magnetic force microscopy, and nitrogen vacancy scanning magnetometry. Skyrmions remain stable at a wide range of applied magnetic fields and are confirmed to be nearly Bloch-type from micromagnetic simulation and analytical magnetization reconstruction. Furthermore, we observe skyrmion pairs which may be explained by skyrmion-antiskyrmion interactions. Our findings expand the family of magnetic materials hosting RT magnetic skyrmions by tuning g-DMI via gradient polarity and choice of magnetic elements.

Magnetic skyrmions are topologically nontrivial vortex-like quasiparticles that possess nano- to-microscale dimensions and high controllability through current induced spin torque.[1] When stabilized at room temperature (RT), they could be used as memory and logic elements promising for next generation energy efficient memory and logic devices[2] as well as neuromorphic computing.[3,4] Dzyaloshinskii Moriya interaction (DMI) is an indirect and anti-symmetric exchange interaction favoring the formation of magnetic skyrmions. Néel skyrmions, Bloch skyrmions, and antiskyrmions have been experimentally observed in various materials. In ultrathin ferromagnetic/heavy metal (FM/HM) multilayers, interfacial DMI[5–7] originates from broken inversion symmetry at interfaces between the FM and HM layers with strong spin orbit coupling (SOC).[1] In this case, Néel-type skyrmions have been identified not only in monolayer Fe/Ir(111) and bilayer PdFe/Ir(111)[8,9] but also in several FM/HM heterostructures composed of sub-nanometer Co layers sandwiched between HM layers (Ir, Pt, W).[7,10–12] Examples include [Ir/Co/Pt]$_{10}$ multilayer structures.[10] However, interfacial DMI strength in FM/HM multilayer systems vanishes by increasing the thickness of the FM layer to a few nanometers and the stability of the magnetic skyrmions relies heavily on the quality of interfaces. In contrast, bulk DMI induced Bloch-type skyrmions in chiral magnet B20-type compounds such as MnSi,[13] Fe$_{1-x}$Co$_x$Si,[14] and FeGe were observed at low temperatures.[15,16] Recently, RT skyrmions were observed in chiral magnet Co$_8$Zn$_{10}$Mn$_2$ that can be manipulated electrically.[17,18] However, materials with bulk DMI are rare and the DMI strength lacks tunability since it is set by the crystal structure. Furthermore, antiskyrmions, characterized by opposite winding number and anisotropic helicity, can be stabilized at RT by anisotropic DMI[19] in Heusler compounds Mn$_{1.4}$PtSn[20,21] and Mn$_{1.4}$Pt$_{0.9}$Pd$_{0.1}$Sn.[22,23] Skyrmion-antiskyrmion pairs were also observed in B20-type FeGe.[16]

Very recently, sizeable DMI was realized at RT by a compositional gradient engineering in single layer films consisting of FM and HM, referred to as gradient DMI (g-DMI). The resulting g-DMI originates from the combined bulk magnetization asymmetry (BMA) and SOC, and has sign and strength dependence with the magnetization gradient.[24,25] Although it is believed that magnetic skyrmions could be stabilized by g-DMI, the direct observation of spin textures in gradient samples is still missing. In addition to g-DMI, non-equilibrium spin torque could originate from the composition gradient induced symmetry breaking. In particular, field-free spin-orbit torque induced magnetization self-switching is reported in gradient CoTb[26] and CoPt[27] single layers. Magnetic skyrmions combined with spin torque extends the application of gradient magnetic single layer systems to spintronics. Motivated by these findings, we have investigated the presence of topological spin textures in gradient CoPt alloy single-layers through direct magnetic imaging.

Here, we use scanning nitrogen-vacancy (NV) magnetometry in combination with magnetic force microscopy (MFM), Brillouin light scattering (BLS), and topological Hall effect (THE) to observe RT magnetic skyrmions in composition gradient engineered CoPt single layers. Isolated skyrmions are measured by MFM and found to correlate to the field dependent topological Hall signal, indicating non-zero topology. Micromagnetic simulation and analytical magnetization reconstructions extracted from NV magnetic stray-field maps reveal nearly Bloch type skyrmions. The isolated skyrmions remain stable at a wide range of applied magnetic fields. Pairs of skyrmions were also observed which may be attributed to higher-order winding number skyrmions, trivial skyrmions, or by skyrmion–antiskyrmion interaction. Of particular interest is observing/controlling exotic topologically protected spin textures in composition gradient magnetic single layer systems and greatly expanding the selection range of materials for the research and application of magnetic skyrmions.[1]

**RESULTS AND DISCUSSION**

**Structural and magnetic properties of CoPt single layer with gradient.** To host topological spin textures, DMI was introduced into binary Co$_x$Pt$_{1-x}$ films with perpendicular magnetic anisotropy (PMA) via a composition gradient induced bulk inversion asymmetry.[24,27] In the framework of three-site Fert-Lévy model, the net bulk DMI is stabilized due to the compositional gradient of FM/HM single layer shown in Figure 1a, where an anti-symmetric exchange interaction between neighboring FM (blue spheres) atoms (sites $i$ and $j$) is enabled by the SOC of the HM (grey spheres) atom (site $l$) via conduction electrons. The DMI vector is given by the formula:[28]

$$\vec{D}_{ijl}(\vec{R}_{li}, \vec{R}_{lj}, \vec{R}_{ij}) = -V_1 \frac{(\vec{R}_{li} \cdot \vec{R}_{lj})(\vec{R}_{li} \times \vec{R}_{lj})}{|\vec{R}_{li}|^3 |\vec{R}_{lj}|^3 |\vec{R}_{ij}|},$$ Eq. 1

where $|\vec{R}_{li}|, |\vec{R}_{lj}|$ and $|\vec{R}_{ij}|$ are the distance vectors and $V_1$ is the SOC material parameter.[25]

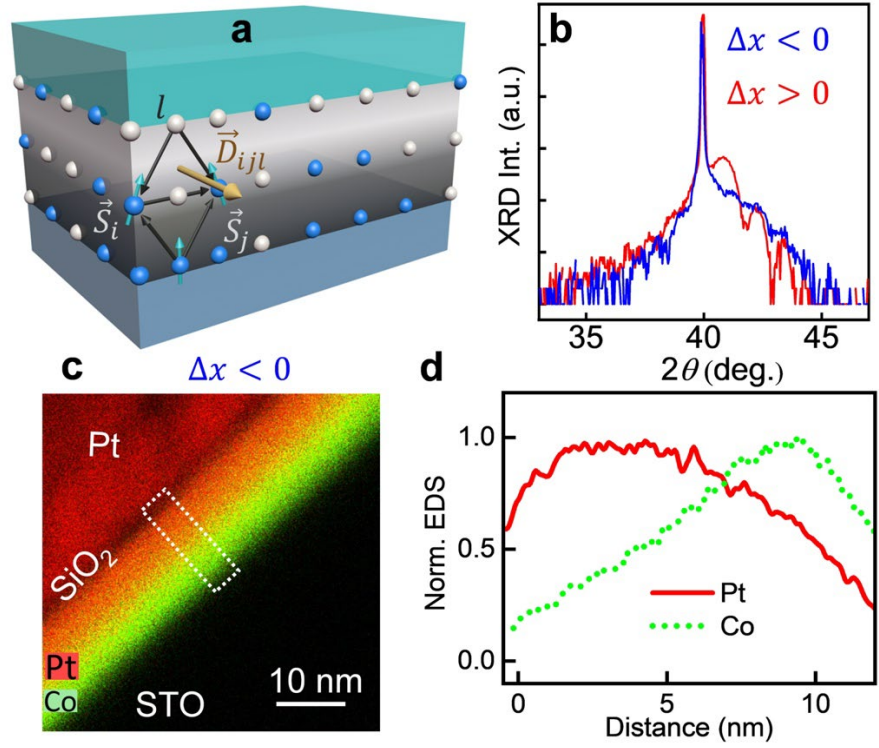

**Figure 1. Structural characterization of g-CoPt single layer with g-DMI.** (a) Schematic of the bulk DMI arising in FM/HM film from the combination of BMA induced by composition gradient and strong SOC in HM. The induced DMI vector $\vec{D}_{ijl}$ can be described by the three site Fert-Lévy model, with magnetic atoms ($\vec{S}_i, \vec{S}_j$) and heavy metal atom ($l$) participating in asymmetric exchange interaction. (b) XRD spectrum for $\Delta x$ = -50% (blue) and $\Delta x$ = +50% (red) g-CoPt single layers. (c) STEM-EDS image of $\Delta x$ = -50% g-CoPt with plotted contribution of Co (green) and Pt (red). (d) Normalized EDS intensity for Pt (red solid line) and Co (dotted green line) for $\Delta x$ = -50% g-CoPt, integrated across the dashed rectangle in (c).

The starting and ending components of $Co_xPt_{1-x}$ defined as $Co_{x_i}Pt_{1-x_i}$ and $Co_{x_t}Pt_{1-x_t}$ give the value of gradient $\Delta x = x_t - x_i$, wherein, $x_i (x_t)$ represents the starting (ending) stoichiometric ratio of Co. 10-nm thick $Co_xPt_{1-x}$ single layer films with positive ($\Delta x = +50\%$) and negative ($\Delta x = -50\%$) composition gradient were synthesized using a co-sputtering technique (see Methods), which is similar with that used in the reference 25. The corresponding composition from start to end is $CoPt_3 \rightarrow Co_3Pt$ ($Co_3Pt \rightarrow CoPt_3$) for positive gradient $\Delta x = +50\%$ (negative gradient $\Delta x = -50\%$). In the following, we refer to gradient $Co_xPt_{1-x}$ single layers as g-CoPt. The smooth surfaces were confirmed through atomic force microscopy (AFM) topography measurements, see Supporting Information (SI) Section S1 and Figure S1.1. The high-resolution x-ray diffraction (HR-XRD) patterns were measured on both $\Delta x = -50\%$ and $\Delta x = +50\%$ g-CoPt films shown in Figure 1b. The (111) diffraction peaks of g-CoPt films with $\Delta x = -50\%$ and $\Delta x = +50\%$ were observed around 42°. The broadening of diffraction peaks may stem from the contribution of $Co_xPt_{1-x}$ possessing different compositions. It is noted that the (111) peak of homogeneous $Co_xPt_{1-x}$ crystal shifts towards higher position with the increase of $x$, owing to the smaller atomic size of Co than that of Pt (see SI Figure S1.2).

Energy dispersive x-ray spectroscopy (EDS) in a scanning transmission electron microscopy (STEM) configuration was used to map the compositional aspects of the $\Delta x = -50\%$ g-CoPt single layer. The EDS compositional map shown in Figure 1c reveals the presence of Co and Pt through the gradient film thickness. It should be noted that the Pt region in the upper left portion of the image is the Pt layer deposited to protect the CoPt film from any damage from Ga ions during STEM sample preparation using standard lift-out techniques.[29] The thin $SiO_2$ layer delineates the protective Pt layer from the film. The relative Co and Pt content in the film (Figure 1d) shows the evolution of the composition through the thickness of the film. This was obtained by quantifying the individually normalized Co and Pt content within the dashed area shown in Figure 1c. For most of the film, the relative EDS signal is linearly varying, suggesting that the magnetization is distributed asymmetrically.

To quantify the magnetism of g-CoPt films, magneto-optical Kerr effect (MOKE), superconducting quantum interference device (SQUID) magnetometry, and BLS spectroscopy were used. Figure 2a shows the normalized polar MOKE signal intensity as function of the applied magnetic field $H_{app}$ for g-CoPt ($\Delta x = -50\%$) and g-CoPt ($\Delta x = +50\%$) films, demonstrating a difference of PMA. $M$–$H$ hysteresis loops measurements with out-of-plane (OOP) and in-plane (IP) applied magnetic field were conducted by SQUID magnetometry (Methods). Figures 2b and 2c show the $M$–$H$ hysteresis loops measured in g-CoPt films with $\Delta x = -50\%$ and $\Delta x = +50\%$. The corresponding saturation magnetization ($M_s$) and effective perpendicular magnetic anisotropy constant ($K_{eff}$) could be obtained: similar $M_s$ values of 717 kA/m for both films, $K_{eff}$ = 0.384 mJ/m$^2$ and 1.11 mJ/m$^2$ for g-CoPt ($\Delta x = -50\%$) and g-CoPt ($\Delta x = +50\%$) films, respectively. The contrast of $K_{eff}$ between negative and positive g-CoPt films may originate from the influence of crystallographic degree under different stacking order (see SI Section S1).

It is predicted that BLS spectroscopy is a powerful technique to quantify DMI. Both the amplitude and sign of DMI could be determined by measuring the nonreciprocal frequency shift of Damon-Eshbach (DE) spin waves.[30–33] The frequency shift $\Delta f$ between the counterpropagating DE spin waves ($\pm k_x$) induced by DMI is given by

$$\Delta f = f_{DM}(-k_x, M_z) - f_{DM}(k_x, M_z) = \frac{2\gamma}{\pi M_s} D k_x, \quad \text{Eq. 2}$$

where $\gamma$, $M_s$, $D$ and $k_x$ are the gyromagnetic ratio, the saturation magnetization, the volume-averaged DMI constant and the projection of the spin-wave vector (**k**) in the direction perpendicular to the applied magnetic field $\mu_0 H$, respectively. Figures 2d and 2e show typical BLS spectra for g-CoPt ($\Delta x = \pm 50\%$) films with $k_x = 18.09$ rad/μm, in which a positive (negative) frequency shift corresponds to the film with positive (negative) gradient. The evolution of $\Delta f$ vs $k_x$ for g-CoPt ($\Delta x = \pm 50\%$) films are summarized in Figure 2f. The opposite slope of $\Delta f$ vs $k_x$ suggests that positive (negative) magnetization gradient gives rise to positive (negative) DMI. Through the linear fitting, the volume-averaged DMI constants $D$ could be obtained: 0.15 mJ/m$^2$ for g-CoPt ($\Delta x = +50\%$) films and -0.51 mJ/m$^2$ for g-CoPt ($\Delta x = -50\%$) films. The difference of DMI amplitude in g-CoPt ($\Delta x = \pm 50\%$) films is observed to which the additional long-range asymmetries along the thickness direction may also contribute.

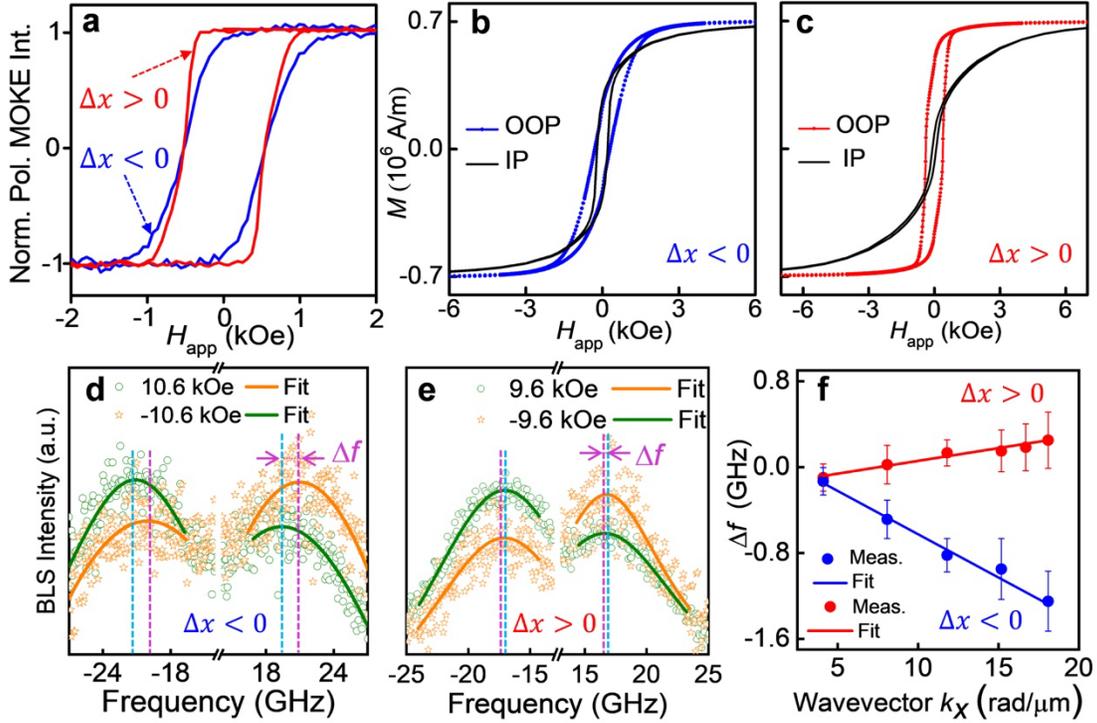

**Figure 2. Magnetic characterization of g-CoPt single layer with g-DMI.** (a) Normalized MOKE intensity vs applied magnetic field for $\Delta x = -50\%$ and $\Delta x = +50\%$ g-CoPt films. OOP (scattered lines) and IP (solid lines) $M$–$H$ hysteresis loops of g-CoPt films with compositional gradient $\Delta x = -50\%$ (b) and $\Delta x = +50\%$ (c). BLS spectra for $\Delta x = -50\%$ (d) and $\Delta x = +50\%$ (e) g-CoPt films in which the scattered green and orange curves represent the spectra with $\pm \mu_0 H$ and the solid lines are the fitting curves. (f) $\Delta f$ vs $k_x$ for g-CoPt ($\Delta x = \pm 50\%$) films.

**Magneto-transport and MFM characterization of g-CoPt single layers.** The magnetic field dependent spin textures are studied on the positive and negative g-CoPt ($\Delta x = \pm 50\%$) films. As electrons pass through magnetic skyrmions, an additional THE signal contributes to the Hall resistance. The relationship between THE signal and magnetic textures with the change of magnetic field could be a fingerprint for the presence of skyrmions.[34] The g-CoPt films were fabricated (see Methods) into Hall bar devices for transport measurement. Then, the Hall resistance $\rho_{xy}$ was measured with a swept OOP magnetic field $H_{app}$. The contribution of longitudinal resistivity could be excluded due to the disappeared anisotropic magnetoresistance and large length-width ratio. To extract the THE signal $\rho_{TH}(H_{app})$,[34] the residual resistivity $\Delta\rho_{xy}(H_{app})$ is estimated through the fit of $\rho_{xy}(H_{app})$ to $\rho_{xy}^{fit}(H_{app}) = R_0 H_{app} + R_s M(H_{app})$ (see SI Section S2). The magnetic hysteresis loops $M(H_{app})$ for both $\Delta x = -50\%$ and $\Delta x = +50\%$ g-CoPt single layers were measured by SQUID (see Figure 3a and Figure S2a). The presence of non-zero $\Delta\rho_{xy}(H_{app})$ is also confirmed when measured by MOKE (Figure 2a), which is collected in the center of the Hall bar (see the inset of Figure 3a). The features of the residual signal $\Delta\rho_{xy}(H_{app})$ between $\rho_{xy}(H_{app})$ and $\rho_{xy}^{fit}(H_{app})$ tracks the accumulation of the additional Berry phase contributed by the itinerant electrons interacting with topological spin textures (*e.g.* skyrmions).[34]

MFM was used to investigate the magnetic domain structure morphology associated with the THE signal $\Delta\rho_{xy}(H_{app})$ in Figure 3a, performed on g-CoPt ($\Delta x = -50\%$) film. The applied magnetic field $H_{app}$ is swept from positive to negative saturation with MFM scans being acquired at intervals throughout the sweep (Figure 3a, I – VIII). Initially in the magnetic field point (I), MFM displays mostly noise level signal with small features, corresponding to the nucleation of skyrmions, correlating with the initial increase in $\Delta\rho_{xy}(H_{app})$. At magnetic field points (II – V), there is a continuation of the nucleated skyrmion structures with decreasing applied magnetic field until the domain structure begins to collapse into the spiral phase, and a more rapid magnetization reversal process takes place. This is reflected in the subsequent increase, maximum, and decrease in the THE signal. As the increasingly negative magnetic field is applied, we observe a somewhat symmetric behavior about -0.3 kOe in both the THE and MFM images (VI – VIII), and the THE signal approaches zero as the film approaches the FM phase. Interestingly, there is even the preservation of certain domains position after passing through the coercive equilibrium, likely associated with pinning centers. A similar comparison between MFM and THE in g-CoPt ($\Delta x = +50\%$) film is shown and discussed in SI Section S2.

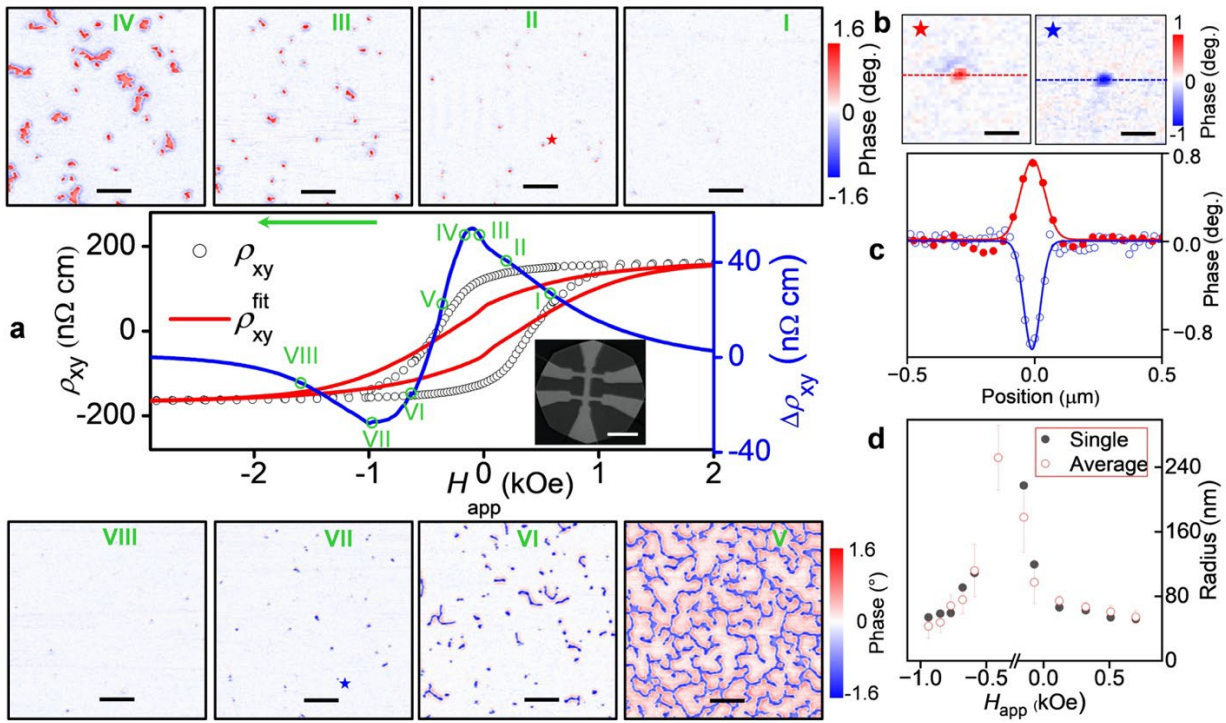

**Figure 3**. **THE and MFM imaging of magnetic morphology through magnetic field sweep in g-CoPt ($\Delta x$ = -50%) single layer.** (a) THE curve (blue line) obtained by subtracting the fitting curve $\rho_{xy}^{fit}$ (red line) from the Hall resistance curve (open circles). Inset of (a): Hall device (scale bar is 50 μm). (I-VIII) Selected MFM images under different magnetic fields indicated by hollow green circles plotted on the THE curve. The scale bar in images I through VIII is 2 μm. (b) MFM images of individual skyrmions (denoted by red and blue stars in images II and VII, respectively) with color coded cross sections (circles) and gaussian profile fitting (solid lines) in (c). (d) Comprehensive radius versus $H_{app}$ obtained from MFM images, in which solid black circles (hollow red circles) represent single skyrmion radius (mean radius).

Details of isolated skyrmions containing opposite core magnetization are shown in Figure 3b. The reversal of magnetic force interaction (opposite phase) from oppositely magnetized skyrmions indicates the preservation of MFM probe magnetization throughout the duration of the applied magnetic field sweep. Information about high coercivity, low moment MFM tips can be found in Methods and reference 35. Line-cuts of the isolated skyrmions profiles are fitted by gaussian functions (Figure 3c). The extracted radius is then plotted versus the applied magnetic field $H_{app}$ (Figure 3d), showing both the average behavior as well as the selected single skyrmion radius. In cases where the shape of isolated skyrmions deviated significantly from radial isotropy, thresholding techniques were employed, and the mean radius of the amorphous area was measured. The error bar represents the interquartile range of the measured skyrmion radii. It is worth noting that the full width at half maximum (FWHM) of the gaussian profile measured by MFM in Figure 3c is still a convolved signal and would require another level of analysis to estimate the radius defined by the underlying magnetization structure of the skyrmion. Furthermore, MFM images are inherently subject to the distortion caused by the back action of the magnetic field produced by the tip on the sample domain structure.[34,36] This is especially problematic in cases where the film has high PMA, where the attractive magnetic field from the tip can more easily cause premature magnetization reversal. This effect was seen when imaging the 10-nm thick g-CoPt ($\Delta x$ = +50%) single layer, preventing the full comparison between THE and MFM (see Figure S2 in SI Section S2). MFM was also performed on the homogeneous $Co_3Pt$ and $CoPt_3$ films as shown in Figure S1-3. The lack of observed spin textures in the ungraded samples is an indication that the magnetization gradient is integral in stabilization of topological spin textures. This correlates well with BLS measurements of close to zero DMI in both $Co_3Pt$ and $CoPt_3$ films, discussed in SI Section S1 and Figure S1.4.

**Vector magnetometry of isolated skyrmions**. Recently, another technique has emerged for measuring magnetic fields at the nanometer scale without interfering with magnetic samples based on optical detection of the electron spin resonances of NV centers in diamond.[37–43] Negatively charged NV centers, composed of a substitutional nitrogen adjacent to a vacancy site (Figure 4a), are bright, stable single photon emitters that exhibit high-contrast optical detected magnetic resonance (ODMR).[44] NV scanning probe microscopy (NV-SPM) has been widely used to quantitatively measure static and dynamic magnetic stray fields of solid-state systems with a good combination of spatial resolution and magnetic sensitivity in a wide range of temperatures (0.3 – 600 K).[45–51] In particular, NV-SPM was used recently to image skyrmions in Co/Pt multilayers,[52,53] CoFeB thin films[54] and Heusler alloys.[55]

In order to gain quantitative insight into the magnetization configurations of the positive and negative gradient composition g-CoPt films, we used NV-SPM (Figure 4b, see Methods and SI Section S3 for further details).[49,56,57] The orientation of the NV atomic structure with respect to the cartesian reference frame is schematized in Figure 4a, indicating polar ($\theta_{NV}$) and azimuthal ($\phi_{NV}$) angles of 53° and 90°, respectively.[58] An external magnetic field $H_{app}$ is applied along the NV symmetry axis to break the degeneracy of the $m_s = \pm 1$ state, creating a pair of spin transitions whose frequencies depend on the amplitude of $H_{app}$.[42,49,57] In addition, the distance between the NV and the g-CoPt surface, $d_{NV}$, is set to find a suitable measurement height which optimizes the spatial resolution without degradation of the ODMR contrast due to spin mixing caused by the strong transverse magnetic stray field components.[59]

Initially, a rapid assessment of the spin textures is obtained using NV photoluminescence (PL) quenching imaging,[48] in which the strong stray field produced by the skyrmions can be read out through its encoded PL signal. PL quenching images of g-CoPt single layer with compositional gradient $\Delta x$ = -50% (Figure 4c) and $\Delta x$ = +50% (Figure 4e) are obtained at $H_{app}$ of 0.1 kOe. This method allows for a detailed map of the size and position of skyrmions in a static domain configuration, as demonstrated in [Pt/Co/Ru]$_{\times 2}$ multilayer stack.[48] Further discretion between the strength regimes of the magnetic stray field can be inferred by the distinction between positive (enhancement) and negative (quenching) photodynamics,[59] see SI Section S6.

Once suitably isolated skyrmions are located, ODMR imaging is performed to measure the parallel component of the stray field, $B_\parallel$, generated by the skyrmions at distance $d_{NV}$ above the magnetic sample. This distance is chosen based on the size of the skyrmion, which affects the resulting stray field magnitude.[52] The Hamiltonian of the system in the cartesian lab coordinates $(x, y, z)$ in Figure 4a is:[44]

$$H = DS_z^2 - \gamma_{NV}(S_x(H_{app,x} + B_x) + S_y(H_{app,y} + B_y) + S_z(H_{app,z} + B_z)), \quad \text{Eq. 3}$$

where $\gamma_{NV} = 28$ GHz/T is the gyromagnetic ratio of the electron spin, $B_x$, $B_y$, and $B_z$ are the cardinal projections of the stray-field vector $\mathbf{B}$. By considering a linear dependence of the components of $B_\parallel$ in the Fourier space (see SI Section S5), one can use the upward propagation protocol to reconstruct the two-dimensional (2D) maps of $B_z$, $B_x$, and $B_y$ at any height above the measurement plane from only one NV measurement of $B_\parallel$.[40,49,57] ODMR images for the isolated skyrmions highlighted in the PL maps in Figure 4c and Figure 4e are shown in Figures 4d and 4f, obtained on compositional gradient -50% and +50% g-CoPt, respectively.

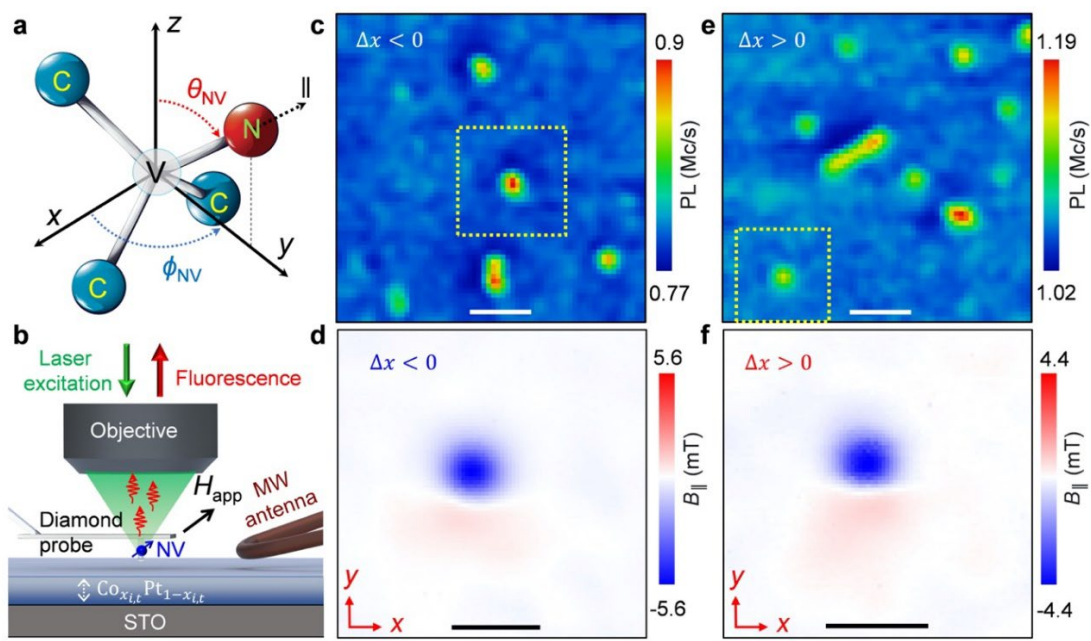

**Figure 4. NV-SPM imaging of spin textures in g-CoPt single layer.** (a) Sketch of the NV molecular structure in the diamond lattice, showing the NV sensing axis parallel (∥) to the applied magnetic field $H_{app}$ relative to cartesian reference frame $(x, y, z)$. (b) Schematic of the NV-SPM imaging apparatus, showing a high NA (0.7) objective used to focus a green (532 nm) laser on the diamond probe with single NV in the confocal geometry. NV PL images of skyrmions in g-CoPt film with compositional gradient $\Delta x$ = -50% (c) and $\Delta x$ = +50% (e), respectively. The scale bar in (c) and (e) is 1 μm. The NV standoff, $d_{NV}$, is 150 nm. (d,f) Two-dimensional $B_\parallel$ magnetic image for the skyrmions denoted by their scan area (yellow dashed square) in (c) and (e). NV standoff is 230 nm and 200 nm in (d) and (f), respectively. The scale bar in (d) and (f) is 500 nm.

Taking the results of g-CoPt ($\Delta x$ = -50%) film for example, Figure 5b shows the simulated 2D $B_\parallel$ stray field of an isolated skyrmion (radius of 120 nm) compared with the measured one (Figure 5a), in which the resulting stray field is calculated by the NV parameters ($d_{NV}$ = 320 nm, $\theta_{NV}$ = 53°, $\phi_{NV}$ = 90°) and the material parameters corresponding to calibration measurements ($M_s$ = 700 kA/m, $K_u$ = 420 kJ/m$^3$) are employed. The value of g-DMI measured by BLS for $\Delta x$ = -50% g-CoPt (-0.5 mJ/m$^2$) was implemented in micromagnetic simulations as a uniform distribution through the thickness of the film, a property of g-DMI which was predicted from atomistic simulations.[24,25] Note that interfacial contribution can induce nonuniformity of total DMI; however, we do not observe any impact of such variation due to small thickness of samples. The extracted line profiles from both measured (Figure 5a) and simulated (Figure 5b) $B_\parallel$ images are plotted in Figure 5c. Of note, the magnetic pattern amplitude $\Delta B$ can be calculated by subtracting the maximum and minimum values of $B_\parallel$.[43] Figure 5d displays $\Delta B$ as function of $d_{NV}$ in the range of 150 nm to 500 nm, deduced from NV measurement (filled circles) and micromagnetic simulation (solid line) $B_\parallel$ images. There is a good agreement between the measured and simulated $\Delta B$ obtained values in the $d_{NV}$ range of 200 – 470 nm. Below 200 nm, the magnetic stray field produced by the skyrmion is higher than the amplitude of $H_{app}$, which leads to a misalignment of the effective magnetic field ($H_{app} + B_\parallel$) along the NV axis, and a decrease in the NV ODMR contrast.[59] This is clearly seen in the NV PL imaging for $d_{NV}$ < 100 nm where a sharp decrease in PL is observed (see SI Section S6), explained by level mixing of NV spin transitions in the excited state induced by the high values of $B_\parallel$ (~ 50 mT)[59] produced by the skyrmions. The calculated 2D maps of $B_y$ (Figure 5e) and $B_z$ (Figure 5f) are demonstrated to gain information about the vector representation of the skyrmion stray field. Figure 5g shows the simulated spin configuration of the skyrmion vectorially, predicting the hybrid type spin texture with a helicity of around $\pi/4$ stronger DMI.

Helicity and polarity are usually used to describe magnetic skyrmions.[60,61] Helicity is defined as the angle of the global rotation around the z-axis. Note that the helicity is $\pi/2$ (zero) for Bloch (Néel) skyrmions. Polarity describes whether the magnetization points in the positive ($p = 1$) or negative ($p = -1$) $z$ direction at the center of the skyrmion. For Bloch and Néel skyrmions, the topological charge and polarity are equal ($Q = p$). Therefore, the difference in helicity distinguishes Bloch and Néel skyrmions from one another. To determine the helicity of the measured skyrmions in g-CoPt single layers, we systematically deduce the phase diagrams from simulations as a function of DMI strength and $H_{app}$ (see SI Section S8, Figure S8.2) and found that skyrmions stabilize with a helicity that is determined by the competition between DMI and the dipolar interactions. For lower values of DMI, the near Bloch type helicity is stabilized.

Along with micromagnetic simulations, analytical reconstruction methods were employed to check the viability of the simulated skyrmions as a possible solution for the measured demagnetization fields from the g-CoPt films. In agreement with micromagnetic simulations, in reconstructed patterns we observe small deviations of helicity from pure Bloch-type skyrmions characterized by the Bloch gauge, i.e., $\nabla \cdot \boldsymbol{m}_{x,y} = 0$; however, due to relatively large $d_{NV}$ we are not able to make quantitative prediction of helicity from the reconstruction procedure. The inverse problem for finding $\boldsymbol{m}$ from the magnetic field maps can be written as:[52]

$$B_\parallel(\boldsymbol{k}, d) = B_z(\boldsymbol{k}, d)(\cos(\theta_{NV}) - i\sin(\theta_{NV})\cos(\phi_k)) \qquad \text{Eq. 4}$$

$$B_z(\boldsymbol{k}, d) = \frac{\mu_0 M_s}{2}(e^{-dk} - e^{-(d+t)k})(-i\cos(\phi_k)m_x(\boldsymbol{k}) - i\sin(\phi_k)m_y(\boldsymbol{k}) + m_z(\boldsymbol{k})), \qquad \text{Eq. 5}$$

where the above equations define the kernels $\alpha_x$, $\alpha_y$, and $\alpha_z$ corresponding to convolutions in $x, y$ plane with $m_x$, $m_y$, and $m_z$, respectively, and $\mathbf{k} = (k\cos(\phi_k), k\sin(\phi_k))$ is the 2D vector in the reciprocal space.

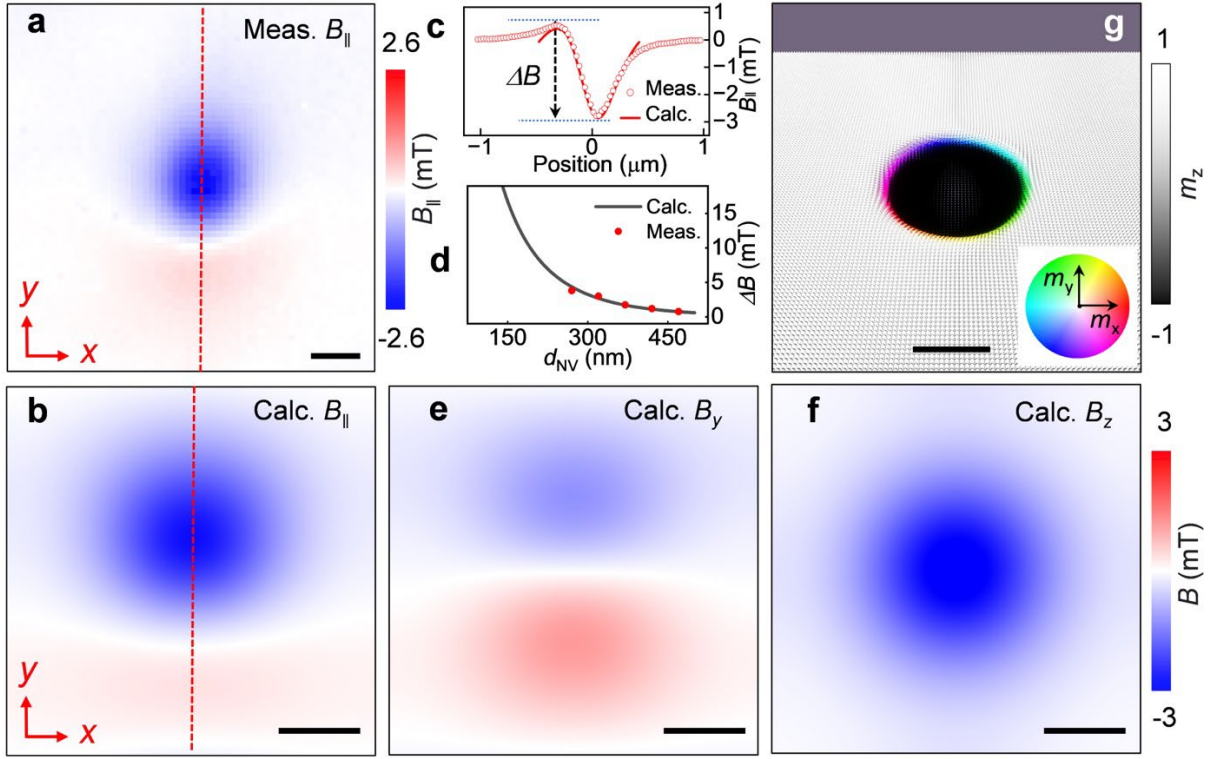

**Figure 5**. **Micromagnetic simulation of isolated skyrmion in g-CoPt ($\Delta x$ = -50%) single layer.** Measured (a) and calculated (b) ODMR $B_\parallel$ image of a single skyrmion from g-CoPt single layer with compositional gradient $\Delta x$ = -50% at an NV standoff $d_{NV}$ of 320 nm and $H_{app}$ of 0.23 kOe. (c) Extracted line profiles from measured (open circles) and simulated (solid line) skyrmion patterns in (a) and (b), respectively. (d) Skyrmion pattern amplitude $\Delta B$ as function of $d_{NV}$ of measured (filled circles) and calculated (solid line) profiles in (c). (e, f) Projections of the stray field at 320 nm above the film surface along the NV sensing axes $y$ and $z$ respectively. The scale bar in (a), (b), (e), and (f) is 200 nm. (g) Vectorial representation of the hybrid skyrmion magnetization. The scale bar in (g) is 100 nm.

A solution for the magnetization vector $\mathbf{m}$ determined up to the null space of the convolution (for example $\mathbf{m}_{x,y}$ of Bloch skyrmion solution corresponds to this null space)[52] is obtained through the implementation of a minimized cost function, achieving deconvolution with respect to the kernel. Combining this reconstruction with the normalization of vector $\mathbf{m}$ and boundary conditions, the magnetization $m_z$ could be reconstructed and the helicity of skyrmions in g-CoPt single layer film can be determined (see Methods). Figure 6 shows the results of the reconstructed magnetization component obtained from the projected demagnetization field for g-CoPt films with $\Delta x$ = -50% and $\Delta x$ = +50%. Figures 6a and 6b show the NV stray field $B_z$ component deduced from the isolated skyrmions in the g-CoPt ($\Delta x$ = -50%) and g-CoPt ($\Delta x$ = +50%) films shown in Figures 4d and 4f, respectively. The $B_z$ component was obtained from NV measured $B_\parallel$ image via a direct transformation made possible due to its linear dependence with the in-plane ($B_x$, $B_y$) components in Fourier space.[40] The corresponding magnetization $m_z$ reconstruction 2D images are plotted in Figures 6c and 6d. Line profiles of the magnetization configurations are shown in Figures 6e and 6f, with cross sections being taken through the skyrmion core center in both *x* and *y* directions on the g-CoPt ($\Delta x$ = -50%) and g-CoPt ($\Delta x$ = +50%) films, respectively.

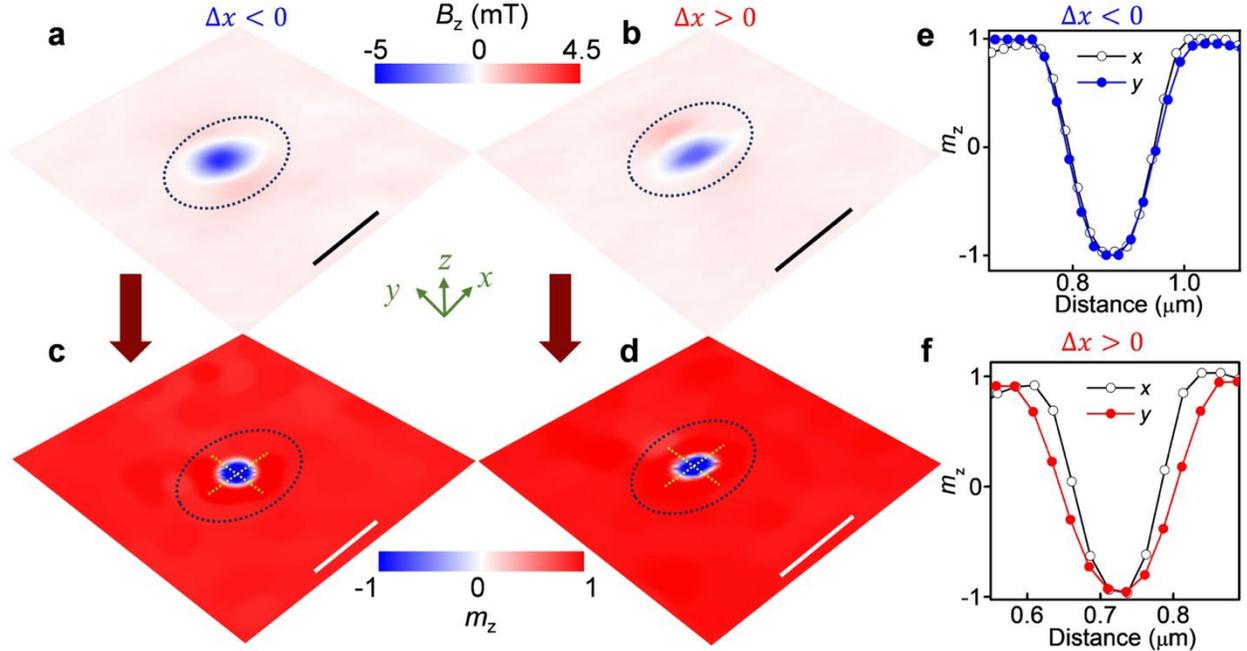

**Figure 6**. **Magnetization reconstruction in the Bloch gauge.** (a,b) $B_z$ maps measured at $H_{app}$ of 0.1 kOe and at a distance $d_{NV}$ = 230 nm (200 nm) above isolated skyrmions on $\Delta x$ = -50% ($\Delta x$ = +50%) g-CoPt film. Subsequent analytical magnetization reconstruction images are shown in (c) and (d) corresponding to $B_z$ images in (a) and (b) respectively. The scale bar in (a–d) is 500 nm. (e,f) Extracted magnetization profiles along *x* (open circle-scattered line) and *y* (filled circle- scattered line) directions from the reconstructed $m_z$ images in (c) and (d).

The resulting normalized magnetization profile for $m_z$ is an indication of the feasibility of the unique solution with small deviations from the Bloch helicity; however, this also required slight adjustments of $d_{NV}$ during the reconstruction as there is uncertainty in experimental determination of this parameter. The reconstruction of the isolated skyrmion in Figure 6d exhibits rotational variance, which may be attributed to pinning sites, and is typical for skyrmions in the low field transitional regime between skyrmion and spiral phase. From the deconvolution procedure, it was also possible to reconstruct components $m_x$ and $m_y$, through the normalization condition $x = |m| = m_x^2 + m_y^2 + m_z^2 = 1$. Furthermore, it is reasonable to assume that the handedness of the stabilized skyrmions depends on the gradient polarity. However, due to the lack of dependence on the $m_x$ and $m_y$ terms in Eq. 5 for Bloch skyrmion, handedness cannot be determined from the measured stray field maps.

In addition to the observation of near Bloch skyrmions in the g-CoPt single layer films, the appearance of skyrmion pairs is reproducibly observed in MFM images for the g-CoPt ($\Delta x$ = -50%) film across a wide range of magnetic fields (up to ± 1.1 kOe), marked by stars in Figures 7a and 7b. There is also the presence of more complicated spin textures (seen in Figure 3a and Figures 7a and 7b) whose morphology and topology may be determined by defects or disorder (see SI Section S8 and Figure S8.1). Similar behavior was observed in Ir/Fe/Co/Pt multilayers, in which wormlike magnetic features were ascribed to be merged Néel skyrmions with topological charge equal to the number of skyrmions.[34] In comparison, the lower density of skyrmions in g-CoPt films would be less likely to form higher order skyrmion chains.[34] To obtain additional information, a zoomed image of a skyrmion pair (marked by green star in Figures 7a and 7b) is shown in Figure 7c, with cross sections along the center-to-center direction, plotted for a sequence of applied fields (Figure 7d). As the amplitude of $H_{app}$ is increased, the individual radii of the pair skyrmions decreases, though the center-to-center distance remains constant, which may be a result of pinning behavior which fixes the domain locally to a structural defect. Importantly, most pairs collapse above a certain applied magnetic field, see the pair highlighted by red and black stars in Figures 7a and 7b. This threshold annihilation behavior with $H_{app}$ may be indicative of a competitive effect between dipolar interactions and magnetic anisotropy to stabilize different features such as skyrmion-antiskyrmion pairs.[16,60,62] Note that we use the term 'skyrmion' to describe the observed (NV and MFM) isolated spin textures in g-CoPt single layers based on the presence of topological charge. A small number of spin textures may also have zero topological charge, and it might be warranted to refer to such textures as magnetic bubbles, see SI Section S8 and Figure S8.1.

A similar occurrence of spontaneously formed pairs is observed by NV-SPM. A large area NV PL map at $H_{app}$ = 0.1 kOe is shown in Figure 7e, in which there are pairs as indicated by a dashed square in addition to isolated spots associated with skyrmions (see SI Section S7). Figure 7f displays spatially resolved PL image of the pairs with a skyrmion distance of 450 nm. Considering that a significant stray-field $B_\parallel$ ~ 15 mT at $d_{NV}$ < 150 nm produced by these skyrmion pairs will significantly reduce NV ODMR contrast (Figure S6.2) related to PL quenching shown in Figure 7f. To measure the amplitude of the stray field $B_\parallel$ along the NV axis, fully quantitative ODMR imaging taken at $d_{NV}$ of 250 nm is performed in Figure 7g, corresponding to the pair imaged in Figure 7f. The skyrmion distance in the pair (450 nm) is higher than $d_{NV}$ of 250 nm, and the individual demagnetization fields can be resolved. As shown in Figure 7g, the $B_\parallel$ amplitude produced by the individual components of the pairs is similar with a slight change of the magnetic pattern, which may be related to the presence of different topological solitons such as skyrmions and antiskyrmions stabilized mainly by dipolar interactions in the g-CoPt single layer.[62] It is known that skyrmions and antiskyrmions can be stabilized by anisotropic DMI or dipolar interactions.[16,60,63] Based on micromagnetic simulations, skyrmion-antiskyrmion pairs as well as skyrmion-skyrmion pairs can be stabilized when pinned by disorder. Figure 7h shows the calculated $B_\parallel$ images for the simulated skyrmion-antiskyrmion pair and it qualitatively agrees with the NV measured $B_\parallel$ map in Figure 7g. Pinned skyrmion-skyrmion pairs can also be stabilized in simulations, as shown in Figure 7i, with a slight change of radius and magnetic stray-field pattern of each skyrmion. The underlying simulated magnetization (Figure S8.3), as well as calculated $B_\parallel$ maps of these systems at lower standoffs $d_{NV}$ (Figure S8.4), are shown in SI Section S8. The stability of higher order topological spin textures (winding number of 2) is also discussed and shown in Figure S8.4. The results of the simulations imply that the inclusion of disorder can influence the resulting magnetic texture in a way which allows for a more diverse set of topological magnetic structures to stabilize. Such spin texture systems may exhibit dynamic properties that are influenced by the local pinning energy landscape, including modification of the skyrmion Hall angle or distortion of the skyrmion morphology upon application of current.[64]

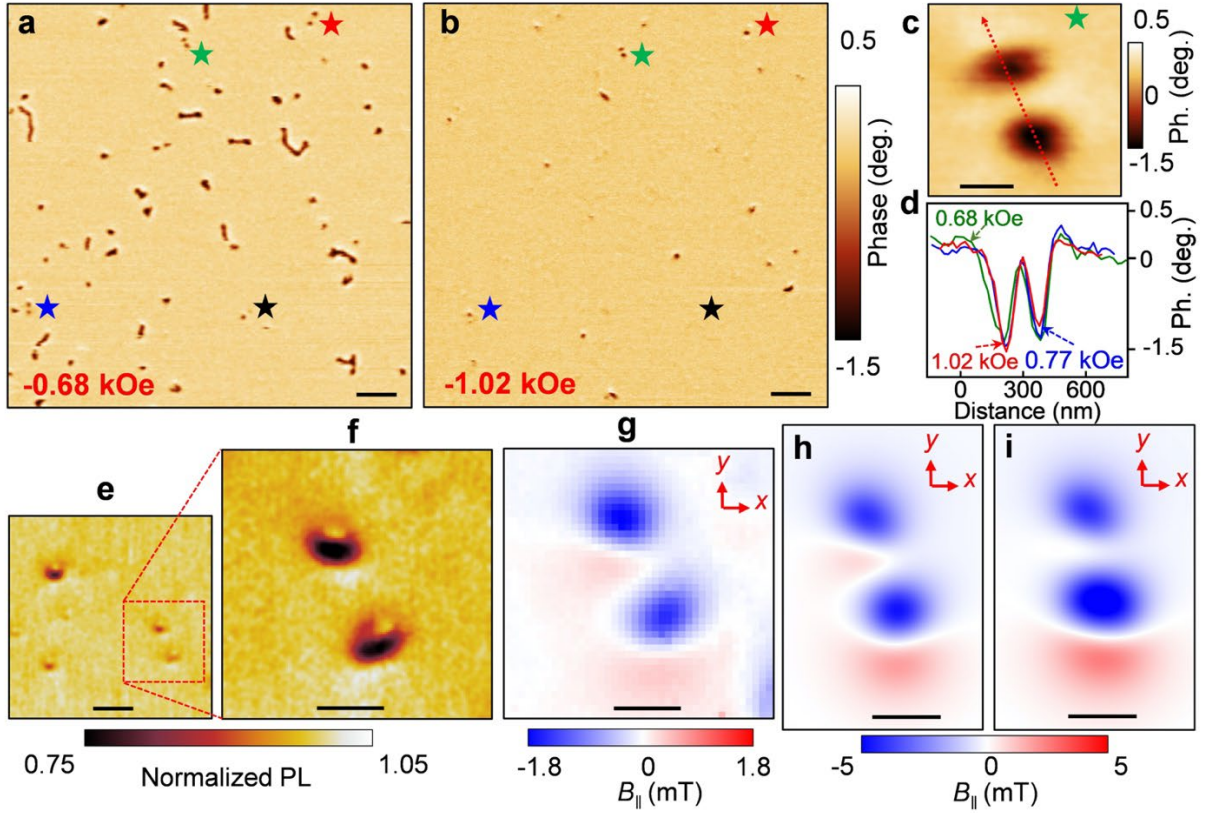

**Figure 7**. **Observation of skyrmion pairs by MFM and NV-SPM in g-CoPt ($\Delta x$ = -50%) single layer.** (a,b) MFM images of skyrmion pairs (filled colored stars) at $H_{app}$ of -0.68 kOe and -1.02 kOe, respectively. The scale bar is 1 μm. (c) Zoomed image of the individual skyrmion pair indicated with a green star in (a) and (b). The scale bar is 100 nm. (d) Line cuts through the pair (red dashed arrow in (c)) for $H_{app}$ of -0.68 kOe, -0.77 kOe, and -1.02 kOe. (e) NV PL quenching image of skyrmions at a standoff $d_{NV}$ of 100 nm and $H_{app}$ of 0.1 kOe. The scale bar is 1 μm. (f) Zoomed image of skyrmion pair indicated by dashed square in (e), taken at $d_{NV}$ of 100 nm below the distance between the pair (~ 450 nm). The scale bar is 500 nm. (g) $B_{\parallel}$ map of the pair in (f), taken at $d_{NV}$ of 250 nm. The scale bar is 500 nm. Simulated magnetic stray field maps for pinned skyrmion-antiskyrmion (h) and skyrmion-skyrmion (i) pairs at a measurement height of 300 nm. The scale bars in (h) and (i) are 500 nm.

## CONCLUSIONS

To summarize, we investigated the spin texture of the topological protected skyrmions in g-CoPt single layers with g-DMI by using NV scanning magnetometry. Skyrmions remain stable at a wide range of applied magnetic fields and are confirmed to be nearly Bloch-type from micromagnetic simulation and analytical magnetization reconstruction profiles extracted from NV magnetic images. The stabilization of pairs is also observed which may be explained by the formation of skyrmion-antiskyrmion[62] or skyrmion-skyrmion[65] pairs, or higher-order skyrmions[34] stabilized by dipolar magnetic interaction and pinned by disorder. Our findings highlight engineered DMI through the gradient composition as an alternative tool to control skyrmions in magnetic materials in a continuous manner.[66] Due to the proportionality between the effective gradient parameter $\Delta x/t$ and its resulting DMI strength, it is possible to further increase the DMI (*i.e.*, through reduction of the thickness for a given gradient amount) and study its effect on skyrmions.[25]

Potentially, in combination with sufficiently low disorder, it may be possible to observe the DMI threshold above which the skyrmions would form a lattice. Furthermore, the ability to tune the PMA and SOC strength by means of gradient sign and choice of elements make gradient DMI in conjunction with NV-SPM an excellent platform for fundamental skyrmionics research.

Other interesting effects arising from the nontrivial topology of skyrmions are their dynamics (skyrmion motion and velocity) in the presence of spin polarized current such as the skyrmion Hall effect.[67] Future magneto-transport experiments combined with NV-SPM[68] by measuring the motion and velocity of skyrmion pairs observed in the g-CoPt films can elucidate the type of these spin textures and their respective velocities as function of the applied magnetic field and film thickness, which may be relevant for application in race-track memories.[2]

## Methods

**Sample preparation.** Stacks with $SiO_2$ (2 nm)/ $Co_xPt_{1-x}$ (10 nm) and a calibration sample Pt (2 nm)/MgO (2 nm)/$Co_3$Pt (4 nm) were deposited by d.c. and radio-frequency magnetron sputtering (Kurt J. Lesker) on a $SrTiO_3$ (STO) (111) single-crystal substrate. During the growth of $Co_xPt_{1-x}$ with composition gradient (g-CoPt), the relative deposition rates of the Co and Pt elements are linearly changed, resulting in a linear composition-magnetization difference along the growth direction. During the growth of the calibration sample $Co_3$Pt, the fixed deposition rates of the Co and Pt elements were employed. The base pressures were lower than $4 \times 10^{-8}$ Torr. During the deposition process of g-CoPt single layers, the temperature was kept at 280 °C and the Ar gas pressure was kept constant at $6 \times 10^{-3}$ Torr. The films were then cooled down to room temperature and a 2 nm $SiO_2$ (MgO) capping layer was deposited by radio-frequency magnetron sputtering to prevent any oxidation effect.

**X-ray diffraction.** X-ray diffraction measurements ($\theta-2\theta$ scan) were performed on positive ($\Delta x = +50\%$) and negative ($\Delta x = -50\%$) composition gradient g-CoPt single layers (thickness of 10 nm) at room temperature at the Singapore Synchrotron Light Source with an X-ray wavelength of 1.541 Å.

**Device Fabrication and Transport Measurement.** The $SiO_2$ (2 nm)/ $Co_xPt_{1-x}$ (10 nm) films with composition gradient $\Delta x = +50\% (-50\%)$ were patterned into a Hall bar device with a width of 5 μm and length of 40 μm by using an Ultraviolet Maskless Lithography machine (TuoTuo Technology) and ion beam etching technology. A calibration sample Pt (2 nm)/MgO (2 nm)/$Co_3$Pt (4 nm)/STO substrate was patterned into a stripe device with a width of 1 μm or 2 μm by electron-beam lithography (EBL) and ion beam etching technology. The measurement of Hall loops was conducted in Physical Property Measurement System (PPMS), in which a dc current of 100 μA was applied and Hall resistance $\rho_{xy}$ can be read with sweeping out-of-plane magnetic field. The magnetic hysteresis loops of un-patterned g-CoPt single layer films with out-of-plane and in-plane magnetic field respectively were measured by SQUID (Quantum design MPMS3).

**Brillouin Light Scattering.** The BLS experiment is under the geometry of Damon-Eshbach (DE) spin-wave modes in all measurements, in which magnetic field $\mu_0 H$ is applied along the film plane and DE spin-wave propagates perpendicularly to the $\mu_0 H$ direction. An s-polarized monochromatic laser beam with wavelength λ = 532 nm is focused onto the sample surface with an incident angle θ. The p-polarized backscattered light was collected and sent to a Sandercock type multipass tandem Fabry–Perot interferometer. Owing to the conservation of momentum on the film plane, the Stokes (anti-Stokes) peak in the BLS spectra corresponds to the creation (annihilation) of magnons with momentum $k_x = \frac{4\pi}{\lambda} \sin\theta$. Due to the presence of the DMI, the counterpropagating DE spin waves became nonreciprocal leading to the frequency shift ($\Delta f$). Then, the amplitude and sign of DMI could be quantified through a linear fitting of $\Delta f$ vs $k_x$ measured at negative and positive $\mu_0 H$.

**Magnetic Force Microscopy.** Magnetic Force Microscope (MFM) images were obtained using a Digital instruments IIIA atomic force microscope. The magnetic tips used were standard Si AFM tips coated with about 20 nm FePt layer and subsequently annealed at 650 °C to achieve a large coercivity of the deposited tip coating and giving relatively low magnetic moment of ~$10^{-14}$ emu at end of tip. The cantilever was magnetized along the out of plane direction under a 3 T magnetic field, ensuring uniaxial magnetization preserves the relative phase contrast during the measurement under the magnetic field. MFM images were taken in double pass mode with a lift height of 30 nm.

**Optically Detected Magnetic Resonance Imaging.** NV stray magnetic field images were obtained using a home-built nitrogen vacancy-scanning probe microscope (NV-SPM) with combined confocal and atomic force microscope functionality.[49,57] Single NV probes were purchased from Qnami with $^{15}$N implantation energy of 12 keV, corresponding to an implantation distance of around 12 nm from the apex of the diamond tip. The negatively charged NV center, composed of a substitutional nitrogen atom adjacent to a vacancy site, is an electronic spin 1 with a spin-triplet ($|m_S = 0\rangle$, $|m_S = \pm 1\rangle$) in the ground state. 532-nm laser illumination induces spin-dependent photoluminescence (650 – 800 nm) allowing optical detected magnetic resonance (ODMR) of its spin state. The applied magnetic field $H_{app}$, provided by a permanent magnet, splits $|m_S = \pm 1\rangle$ state via Zeeman effect and leads to two ($|m_S = 0\rangle$ to $|m_S = -1\rangle$ and $|m_S = 0\rangle$ to $|m_S = +1\rangle$) ODMR peaks whose frequencies depend on the projection of the field along the NV symmetry axis. During ODMR imaging, the $|m_S = 0\rangle$ to $|m_S = +1\rangle$ spin resonance spectrum is obtained at each pixel of the sample area and the Zeeman splitting is measured, creating an image of the resulting Lorentzian center frequency shift. The setup is integrated with microwave (MW) equipment to monitor NV spin transitions and with a single photon counter module (SPCM) coupled with a single mode fiber.[37,49,56] The objective used was a 6 mm WD, 0.7 NA Mitutoyo objective, which allowed for flexibility of setup considering the tip and MW antenna placement. ANP-101 nano-positioners from Attocube were used for course positioning of the tip during approach. The sample is rastered by a closed loop three axis piezo stage (NPXY100Z10-128), with the diamond tip fixed.

**Micromagnetic Simulations.** To model the magnetic skyrmions measured by NV-SPM, we performed micromagnetic simulations using the GPU-based platform Mumax3.[69] The simulations were carried out using the magnetic parameters ($M_s$, $K_u$) of the negative $\Delta x = -50\%$ and positive $\Delta x = +50\%$ g-CoPt films (thickness of 10 nm), obtained from SQUID measurements. The position dependent DMI was included phenomenologically using a function,[24] $C + D_b e^{-(z-z_b)/l} + D_t e^{-(z-z_t)/l}$, where $D_b$ and $D_t$ are the DMI amplitudes at the bottom and top interfaces of the CoPt film with coordinates $z_b$ and $z_t$, respectively, which decay into the film interior (bulk) with a decay length $l$. The DMI contribution from the gradient induced BMA, *C is* kept constant through the film thickness.[24] Periodic boundary conditions were imposed when simulating single skyrmions to prevent demagnetization effects at the edge. Key parameters to fit the NV and micromagnetic simulations are the NV standoff $d_{NV}$, NV polar ($\theta_{NV}$) and azimuthal ($\phi_{NV}$) angles relative to cartesian reference frame (*x*, *y*, *z*), Figure 4a. To measure these parameters, we used a calibration FM substrate composed of 1 μm wide Pt (2 nm)/MgO (2 nm)/Co$_3$Pt (4 nm)/STO stripes with high PMA. The fabrication details and magnetic properties of this sample are provided in the SI Section S4.

**Magnetization Reconstruction of Skyrmion Profiles from NV ODMR Images.** We consider magnetic stray fields produced by magnetic dipoles confined within a magnetic film of thickness $t$. We disregard variations in $M_s$ in the film since it has a small effect for $t \ll d_{NV}$. The stray fields are related to magnetization direction by relation: $\boldsymbol{B}(\boldsymbol{k}, d) = \widehat{D}(\boldsymbol{k}, d)\boldsymbol{m}(\boldsymbol{k})$, where $\boldsymbol{m}(\boldsymbol{k})$ is the 2D Fourier transform of magnetization direction vector and $\boldsymbol{B}(\boldsymbol{k}, d)$ is the 2D Fourier transform of the magnetic field in the plane that is distance $d$ away from the film. The kernel matrix $\widehat{D}(\boldsymbol{k}, d)$ reads as:

$$\widehat{D}(\boldsymbol{k}, d) = \frac{\mu_0 M_s}{2} (e^{-dk} - e^{-(d+t)k}) \begin{pmatrix} -\cos^2(\phi_k) & -\frac{\sin(2\phi_k)}{2} & -i\cos(\phi_k) \\ -\frac{\sin(2\phi_k)}{2} & -\sin^2(\phi_k) & -i\sin(\phi_k) \\ -i\cos(\phi_k) & -i\sin(\phi_k) & 1 \end{pmatrix}, \quad \text{Eq. M1}$$

where the rows of matrix $\widehat{D}(\boldsymbol{k}, d)$ are linearly dependent. We rewrite Eq. M1 in real space as:

$$B_\parallel(\boldsymbol{\rho}) = \frac{\mu_0 M_s}{2} \left( \alpha_x * m_x(\boldsymbol{\rho}) + \alpha_y * m_y(\boldsymbol{\rho}) + \alpha_z * m_z(\boldsymbol{\rho}) \right), \quad \text{Eq. M2}$$

where the symbol $*$ denotes the convolution, $\boldsymbol{\rho}$ defines a two-dimensional position within the reconstruction plane, and we define three convolution kernels $\alpha_x$, $\alpha_y$, and $\alpha_z$. The analytical expressions for kernels can be easily calculated but are complicated and we do not present them here. The solution of Eq. M2 for $\boldsymbol{m}(\boldsymbol{\rho})$ is not unique due to the nonvanishing null space of the kernels. The procedure was similar to reference 52. Nevertheless, reference 52 argued that enforcing normalization and skyrmion boundary conditions at $\rho \to \infty$ selects one solution within the Bloch or Néel gauge. Here, a different cost minimization technique is utilized, which relied on the built-in function ImageDeconvolve in Wolfram Mathematica (https://www.wolfram.com/). Furthermore, reference 52 considered solutions deviating from the above gauges. In our heuristic approach, we directly minimize the following functional:

$$\mathcal{L} = \int \left( B_\parallel(\boldsymbol{\rho}) - \frac{\mu_0 M_s}{2} \vec{\alpha} \cdot \boldsymbol{m}(\boldsymbol{\rho}) \right)^2 d\boldsymbol{\rho}, \quad \text{Eq. M3}$$

by using the built-in function ImageDeconvolve in Wolfram Mathematica. We observe that with enforced skyrmion boundary conditions the minimization procedure chooses solutions that are very close to the Bloch type solutions, which also agrees with our micromagnetic analysis (see the discussion above and in the SI Section S8).

**ASSOCIATED CONTENT**

**Supporting Information**
The Supporting Information is available free of charge at https://pubs.acs.org/

> Characterization (AFM, XRD) of g-CoPt single layer films; Magnetic (THE, SQUID, MFM) characterization of g-CoPt ($\Delta x = +50\%$) film; quantitative NV ODMR imaging; calibration of NV measurements: measuring the standoff and NV axis angles; vector reconstruction of skyrmion stray field components; height dependence in PL quenching imaging; corroboration of MFM and NV quenching imaging; and micromagnetic simulations.

# AUTHOR INFORMATION

**Corresponding Authors**

Abdelghani Laraoui − Department of Physics and Astronomy and the Nebraska Center for Materials and Nanoscience, University of Nebraska-Lincoln, Lincoln, Nebraska 68588, United States; Department of Mechanical & Materials Engineering, University of Nebraska-Lincoln, Lincoln, Nebraska 68588, United States; orcid.org/0000-0002-2811-8030; Email: alaraoui2@unl.edu

Jingsheng Chen − Department of Materials Science and Engineering, National University of Singapore, Singapore 117579, Singapore; National University of Singapore (Suzhou) Research Institute, Suzhou, Jiangsu, 215123, China; orcid.org/0000-0003-3188-2803; Email: msecj@nus.edu.sg

**Author Contributions**

A.E. performed NV-SPM and MFM measurements and analyzed the data; Q.Z, grew the g-CoPt films and performed topography (AFM), XRD, MOKE, SQUID, AHE, and THE measurements; H.V., E.S., and A.A.K. performed micromagnetic modeling and magnetization reconstruction analysis; C.L. and G.C. performed BLS measurements; S.S. and J.E.S. performed HAADF-STEM and EDS on the graded CoPt films; S.L. and S-Y.L. assisted A.E in MFM measurements; L.J. helped to prepare the CoPt stripe device by EBL; I.F. created code for recovering vector stray-field components and analyzed NV-ODMR images. J.C. and A.L. designed the experiments and supervised the project. A.E. and A.L. wrote the paper with contributions and feedback from all authors.

**Notes**

The authors declare that they have no competing financial interests.

# ACKNOWLEDGMENTS

A.L., A.A.K., and S.-H.L. acknowledge support by the National Science Foundation EPSCoR RII Track-1: Emergent Quantum Materials and Technologies (EQUATE), Award OIA-2044049. The research done at University of Nebraska-Lincoln was performed in part in the Nebraska Nanoscale Facility: National Nanotechnology Coordinated Infrastructure and the Nebraska Center for Materials and Nanoscience (and/or NERCF), supported by the National Science Foundation ECCS under Award 2025298, and the Nebraska Research Initiative. The research performed in the National University of Singapore was supported by the Singapore Ministry of Education MOE-T2EP50223-0006, MOE-T2EP50123-0012, MOE-T2EP50121-0011, and MOE Tier 1: 22-4888-A0001. I. F. acknowledges support from the Latvian Quantum Initiative under European Union Recovery and Resilience Facility project no. 2.3.1.1.i.0/1/22/I/CFLA/001.

**Supporting Information**
**S1. Characterization of graded and ungraded CoPt single layer films**

Throughout the experimental duration, the surface topography underwent frequent characterization. Post-growth evaluations were conducted using atomic force microscopy (AFM) on both negative $\Delta x = -50\%$ (Figure S1.1a) and positive $\Delta x = +50\%$ (Figure S1.1b) gradient CoPt (g-CoPt) single layers (thickness of 10 nm), grown on $SrTiO_3$ (STO) substrate. Transverse line profiles (highlighted by dashed lines) are plotted in Figures S1.1c and S1.1d, corresponding to topography images in Figures S1.1a and S1.1b, respectively. Root mean square (RMS) roughness is measured as 110 pm and 234 pm on negative and positive g-CoPt, respectively. In every instance of subsequent topography imaging, the surface displayed sub-nanometer roughness. Moreover, there are no discernible correlations between the magnetic features studied in the main manuscript and their underlying topography, as measured by the height channels of nitrogen vacancy-scanning probe microscopy (NV-SPM) and magnetic force microscopy (MFM).

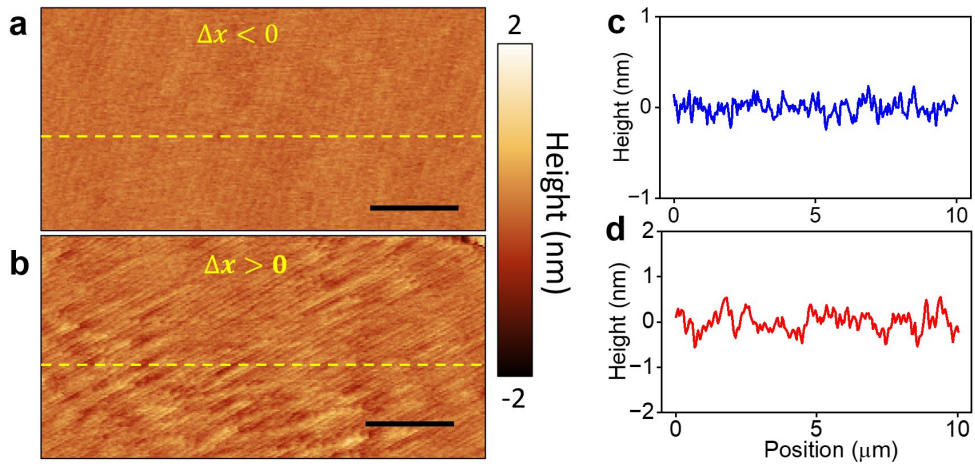

**Figure S1.1. Surface roughness characterization.** Topography image of $\Delta x = -50\%$ (a) and $\Delta x = +50\%$ (b) g-CoPt film. The scale bar in (a-b) is 2 μm. Extracted transverse line cuts for $\Delta x = -50\%$ (c) and $\Delta x = +50\%$ (d) g-CoPt film, highlighted by a dashed line in topography maps in (a,b).

The high-resolution X-ray diffraction (HR-XRD) patterns were measured on both gradient CoPt films and homogeneous $Co_xPt_{1-x}$ films. The (111) peaks of homogeneous $CoPt_3$ and $Co_3Pt$ film are at 40.1° to 43.3° as shown in Figure S1.2e. As $x$ increases from 25% to 75%, the (111) peak shifts from 40.1° to 43.3°, indicating the change of grain size.[27,70,71] The HR-XRD patterns were measured on $\Delta x = -50\%$ and $\Delta x = +50\%$ g-CoPt films shown in Figure 1b and Figure 1c, respectively. The (111) diffraction peaks are around 42° with broadening for both $\Delta x = -50\%$ and $\Delta x = +50\%$ g-CoPt films, providing indirect evidence of the existence of composition gradient. The (111) peak is also related to the origin of PMA.

The magnetic properties of the g-CoPt single layer films were measured by superconducting quantum interference device (SQUID) magnetometer (Quantum design MPMS3). The magnetic hysteresis loops of the un-patterned g-CoPt films with out-of-plane (OOP) and in-plane (IP) magnetic field were measured to obtain anisotropy constant. Then the effective anisotropy constant ($K_{eff}$) is calculated from $K_{eff} = M_s \times \Delta/\mu_0$,[72] where $\Delta$ is the difference in area between the OOP and IP magnetic hysteresis loops, $M_s$ is the saturation magnetization and $\mu_0$ is the permeability of free space. $M_s$ has similar values of 717 kA/m for both $\Delta x = -50\%$ and $\Delta x = +50\%$ g-CoPt films. The corresponding effective perpendicular magnetic anisotropy constant ($K_{eff}$) could be obtained: $K_{eff}$ = 0.384 mJ/m$^2$ and 1.11 mJ/m$^2$ for $\Delta x = -50\%$ and $\Delta x = +50\%$ g-CoPt films, respectively. The $K_{eff}$ of positive gradient (CoPt$_3$→Co$_3$Pt) sample is higher than the negative gradient (Co$_3$Pt→CoPt$_3$) sample, which may be explained by the different crystallographic degree in films with opposite stack order.

In general, the Co$_x$Pt$_{1-x}$ alloys have many crystallographic phases,[73,74] including disordered structures of A1 (cubic, Figure S1.2a) and A3 (hexagonal, Figure S1.2b), ordered structures of L1$_0$ (tetragonal) and L1$_2$ (cubic), and metastable L1$_1$ ordered structures of (rhombohedral, Figure S1.2c) and B$_h$ (hexagonal, Figure S1.2d). Since the close-packed plane usually has a low surface free energy,[75] the easy magnetization axis will be normal to the close-packed plane resulting in the strong perpendicular magnetic anisotropy (PMA). Among all phases, close-packed stackings such as L1$_1$ ordered phase, or hexagonal (hcp) phase show PMA,[76] which has advantages in application of spintronics. Therefore, the magnitude of the anisotropy constant is strongly affected by the degree of chemical order. Of note, the lattice mismatch with the substrate is an important source of defects lowering the degree of chemical order. Considering the hcp structure, the lattice constant ($a = b$) is 5.19 Å, 5.472 Å and 5.529 Å for Co$_3$Pt, CoPt$_3$ and STO substrate respectively. The lattice mismatch between CoPt$_3$ and STO substrate is less than Co$_3$Pt. For positive gradient sample, an initial grown CoPt$_3$ layer could be regarded as the seed layer to help form hcp phase of Co$_x$Pt$_{1-x}$ alloy. Then in-plane Co-Co bonds in the alternating layers of Co and Pt result in PMA. As the film gets thicker, the increase of hcp phase order degree gives rise to the higher $K_{eff}$. For the negative gradient sample, it is predictable that the relatively low order degree of Co$_3$Pt induces a drop in the $K_{eff}$.

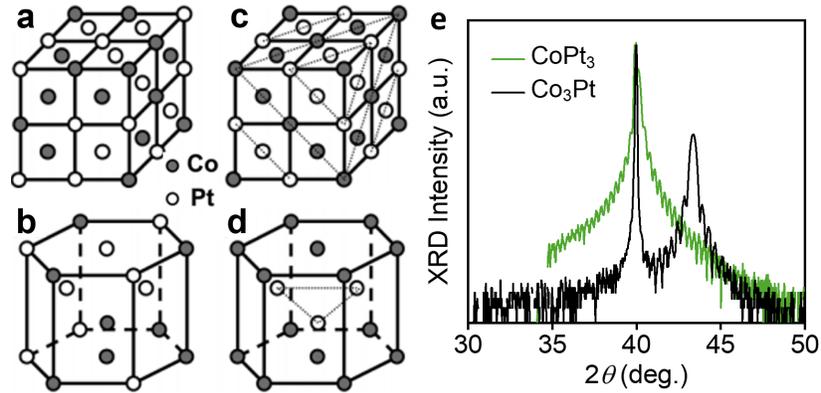

**Figure S1.2. Preferred crystal structures for CoPt single crystal growth.** (a) A1, (b) A3, (c) L1$_1$, and (d) B$_h$ crystal structures. (e) XRD spectrums of CoPt$_3$ and Co$_3$Pt film.

To compare the magnetic properties to the g-CoPt films, we grew a set of Co$_3$Pt and CoPt$_3$ non graded films with a thickness of 10 nm and studied their magnetic properties using SQUID magnetometry, MFM, and Brillouin Light Scattering (BLS). The IP and OOP hysteresis loops of the homogeneous films are shown in Figure S1.3a for Co$_3$Pt and in Figure S1.3b for CoPt$_3$. The Co$_3$Pt film shows PMA, which is in accordance with the structurally induced PMA seen in $\Delta x = +50\%$ g-CoPt film. The CoPt$_3$ film shows in-plane magnetic anisotropy with weak in-plane coercive field of around 50 Oe. Figure S1.3c (I-IV) displays the MFM images of Co$_3$Pt film at different applied magnetic field values (highlighted in Figure S1.3a by I-IV) with a sudden formation of the spiral phase, shrinking of the oppositely magnetized domains, and subsequent saturation. There is a small signal in the MFM images I and IV which is due to a topographic crosstalk from few-nanometer sized bumps on the surface. Due to the CoPt$_3$ sample having in-plane anisotropy, it was not possible to observe the magnetization reversal process with our MFM apparatus. Thus, for several applied out-of-plane magnetic fields (highlighted by letters i-iv in Figure S1.3b), MFM images in Figure S1.3d (i-iv) show zero phase contrast, corresponding to the saturated magnetic state of the CoPt$_3$ film.

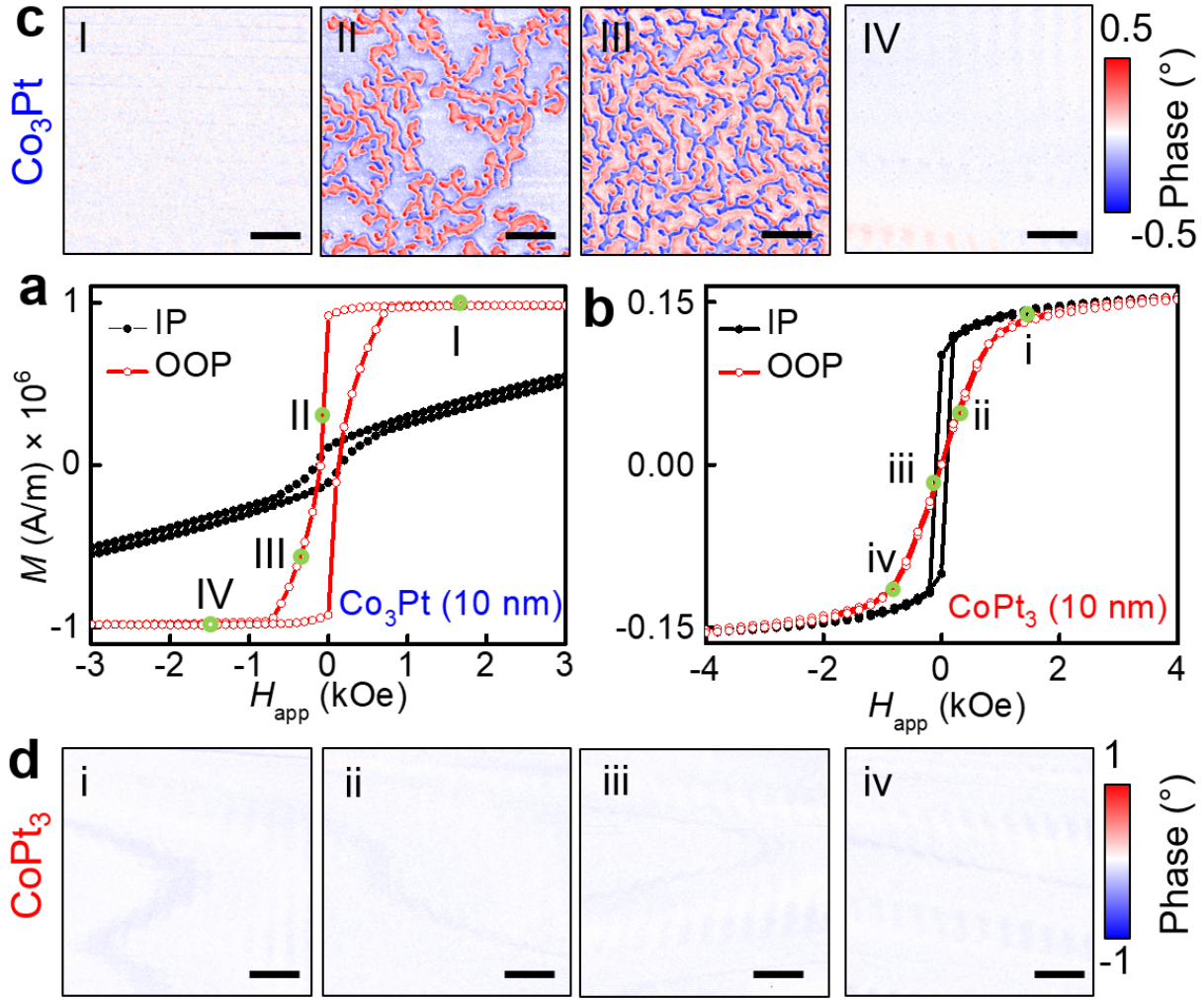

**Figure S1.3.** (a) IP (scattered filled-circle line) and OOP (scattered open-circle line) *M-H* loops of 10 nm thick Co$_3$Pt film. (b) IP (scattered filled-circle line) and OOP (scattered open-circle line) *M-H* loops of 10 nm thick CoPt$_3$. (c) Corresponding MFM images taken at applied magnetic field points I, II, III, and IV in (a). (d) Corresponding MFM images taken at applied magnetic field points i, ii, iii, and iv in (b).

Figures S1.4a and S1.4b show representative BLS spectra for Co$_3$Pt and CoPt$_3$ films, respectively, with $k_x = 18.09$ rad/μm. The evolutions of $\Delta f$ vs $k_x$ for both films are summarized in Figure S1.4c. Linear fitting of the curves lead to a volume-averaged DMI constants *D* of -0.055 mJ/m$^2$ for Co$_3$Pt film and +0.008 mJ/m$^2$ for CoPt$_3$ film. The DMI is negligible in comparison to the graded films as discussed in the main text and in good agreement with MFM results in Figure S1.3 for the absence of any topological spin textures.

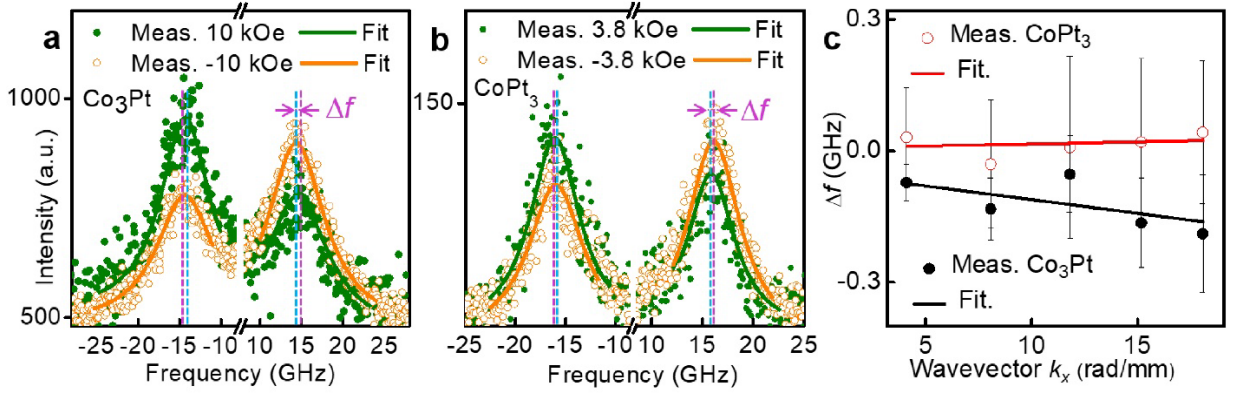

**Figure S1.4.** BLS spectra for $Co_3Pt$ (a) and $CoPt_3$ (b) films in which the green (orange) curve represents the spectra with $\mu_0 H$ ($-\mu_0 H$) and the solid lines are the fitting curves. (c) $\Delta f$ vs $k_x$ for $Co_3Pt$ (filled circles) and $CoPt_3$ (open circles) films.

## S2. Magnetic characterization of g-CoPt ($\Delta x = +50\%$) film

The presence of skyrmions contributes to an additional topological hall effect (THE) signal $\rho_{TH}(H_{app})$ in the Hall resistance $\rho_{xy}$. In order to study the evolution of magnetic textures, the THE signal $\rho_{TH}(H_{app})$ was extracted. This contribution can be quantified by:[34]

$$\rho_{xy}(H_{app}) = R_0 H_{app} + R_s M(H_{app}) + \rho_{TH}(H_{app}),  \qquad \text{Eq. S1}$$

where $R_0 H_{app}$ and $R_s M(H_{app})$ are the ordinary and anomalous Hall components, and $\rho_{TH}(H)$ is the resistivity from the contribution of THE. To extract the THE signal $\rho_{TH}(H_{app})$, the residual resistivity $\Delta\rho_{xy}(H_{app})$ is estimated through the fit of $\rho_{xy}(H_{app})$ to $\rho_{xy}^{fit}(H_{app}) = R_0 H_{app} + R_s M(H_{app})$. The magnetic hysteresis loops $M(H_{app})$ for g-CoPt single layers were measured by SQUID. Besides, the presence of non-zero $\Delta\rho_{xy}(H_{app})$ is also confirmed via using $M(H_{app})$ when measured by MOKE, which is collected in the center of the Hall bar.

Characterization by THE and MFM measurements were also conducted on 10 nm g-CoPt ($\Delta x = +50\%$) film, see Figure S2a. Size dependence of the mean radius as function of the applied magnetic field $H_{app}$ is shown in Figure S2b, with the IQR plotted as error bars. Considering independence due to symmetry, the increase in the PMA, as discussed in the main text, is the main differentiating factor between the positive and negative gradient film properties. Because of the increased PMA, the film was susceptible to unwanted magnetization reversal due to the MFM tip field. This is shown in Figure S2c, with the film being switched midway through the scan (indicated by black star) and proceeding to switch neighboring regions in a cascade process. The same region is subsequently imaged, revealing the stabilized domain structure as well as initiating the same switching process in the previously saturated area. The scan directions are indicated by the green arrows. Such interaction would only impact the scans for a particular alignment of the sample magnetization with respect to the tip magnetization, when the tip and sample magnetization are antiparallel, but the applied magnetic field is approaching the coercive field. As a result, the nucleation field range ($\sim -0.4 - -1$ kOe) could not be imaged without the switching behavior caused by tip-sample interactions.

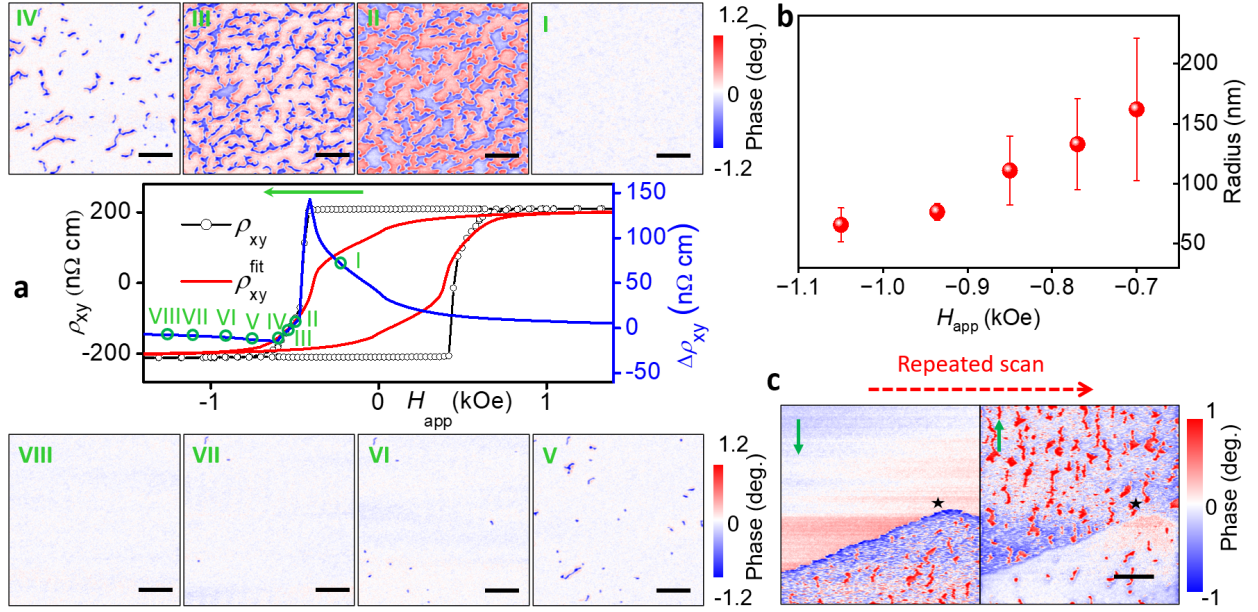

**Figure S2. MFM imaging and THE in g-CoPt (Δx = +50%) single layer.** (a) THE curve (blue line) obtained by subtracting the fitting curve $\rho_{xy}^{fit}$ (red line) from the Hall resistance curve (open circles). I-VIII in (a): Selected MFM images under different magnetic field indicated by open green circles plotted on the THE curve. The scale bar in images I-VIII is 2 μm. MFM lift height was 30 nm. (b) Skyrmion size dependence as function of magnetic field $H_{app}$, with error bars representing the interquartile range. (c) Demonstration of magnetization switching cascade initiated by the magnetic field from the MFM tip. The images are taken consecutively in the same area and at the same applied magnetic field. Green arrows in (c) represent the scan direction. The scale bar in (c) is 2 μm.

### S3. Quantitative NV ODMR Imaging

Quantitative NV stray magnetic field images were obtained using a home-built NV-scanning probe microscope (NV-SPM), as detailed in reference [49]. Figure S3a displays a photoluminescence (PL) image of single NV center in the diamond probe. Figure S3b shows optically detected magnetic resonance (ODMR) peaks for $|m_S = 0\rangle$ to $|m_S = +1\rangle$ NV spin resonance. ODMR maps were obtained by tracking this peak's center frequency while scanning over an isolated skyrmion or skyrmion pair. To optimize the accuracy for a given measurement, the measurements were averaged 10 – 15 times per pixel. Oscillation amplitude of the NV probe was minimized to reduce the positional variation of the sensor. Even at distances of 200 nm above the sample surface, stray field could reach ~ 6 mT. As such, the sweep range was increased to around 60 – 80 MHz to frame ODMR as the field varied rapidly from pixel to pixel. The optimized minimum measurable DC magnetic field in the photon-shot-noise limit is given by:[43,49] $B_{min} \cong 4\,\Gamma\,(3\sqrt{3}\,\gamma_{NV}\,C)^{-1}\,(I_0 t)^{-1}$, where $\Gamma$ is the full-width-at-half-maximum linewidth of the ODMR peak, $C$ is the ODMR peak contrast, $I_0$ is the NV PL rate, and $t$ is the measurements time.[77] By using the parameters of the NV measurements in Figure S3, ($I_0$ = 500k counts/s, $\Gamma$ = 7.78 MHz, C = 0.15) we find $B_{min}$ = 5 μT for $t$ = 1s.

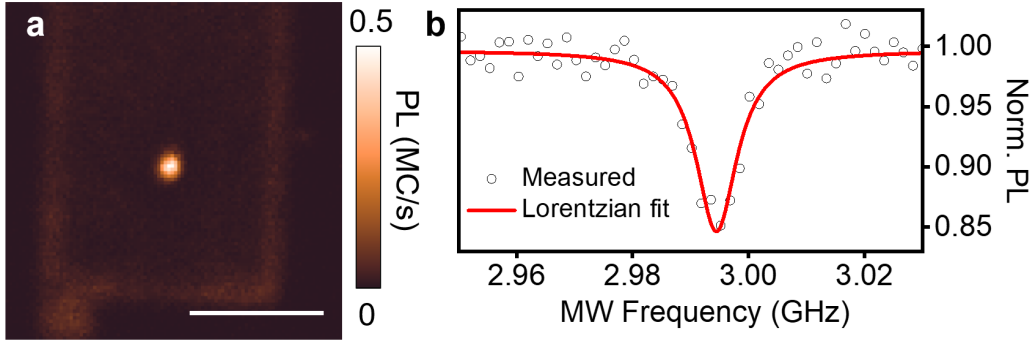

**Figure S3. Characterization of implanted NV center in a diamond probe**. (a) Scanning PL image of NV probe with an implanted single NV. The scale bar is 2 μm. (b) Representative ODMR spectrum of $|m_S = 0\rangle$ to $|m_S = +1\rangle$ peak fitted by a Lorentzian function.

## S4. Calibration of NV measurements: measuring the standoff and NV axis angles

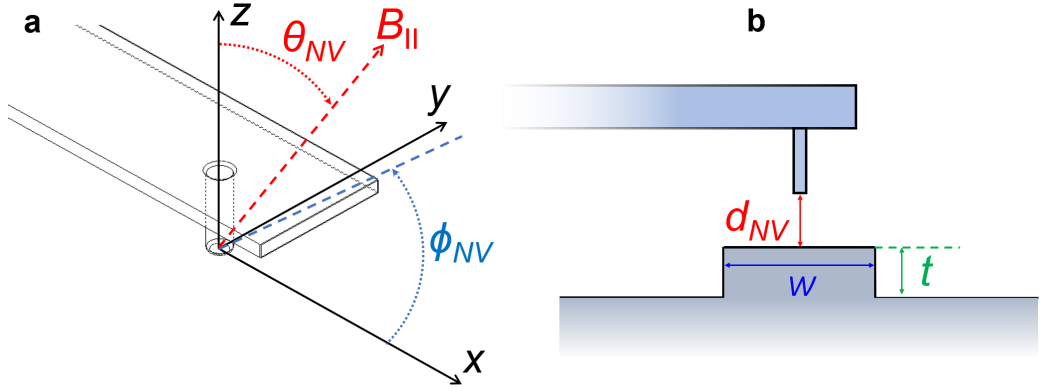

**Figure S4.1. Schematic and parameters for calibration of NV standoff and polar coordinates.** (a) Laboratory reference frame for NV measurements. The NV axis is defined by a polar angle, $\theta_{NV}$, and azimuthal angle, $\phi_{NV}$, measured from the $x$ axis and the $z$ axis, respectively. (b) Wires are described by their width, $w$ and thickness, $t$ and measured at a stand-off, $d_{NV}$.

For an infinitely long wire, oriented along the $y$-axis of a ferromagnetic material with high PMA, the resulting magnetic stray field at the edge of the wire is given by:[78,79]

$$B_x^{\text{edge}}(x, d_{NV}) = \frac{\mu_0 M_s}{4\pi} \ln\left(\frac{x^2 + \left(d_{NV} + \frac{t}{2}\right)^2}{x^2 + \left(d_{NV} - \frac{t}{2}\right)^2}\right) \quad \text{Eq. S2}$$

$$B_z^{\text{edge}}(x, d_{NV}) = \frac{\mu_0 M_s}{2\pi}\left[\text{atan}\left(\frac{x}{d_{NV} + \frac{t}{2}}\right) - \text{atan}\left(\frac{x}{d_{NV} - \frac{t}{2}}\right)\right], \quad \text{Eq. S3}$$

where the edge is defined at $x = 0$, $t$ is the wire thickness, and $d_{NV}$ is the NV sensor-to-sample distance. For simplification in the case of $d_{NV} \gg t$

$$B_x^{\text{edge}}(x, d_{NV}) = \frac{\mu_0 M_s t}{2\pi} \frac{d_{NV}}{x^2 + d_{NV}^2} \quad \text{Eq. S4}$$

$$B_z^{\text{edge}}(x, d_{NV}) = -\frac{\mu_0 M_s t}{2\pi} \frac{x}{x^2 + d_{NV}^2} \quad \text{Eq. S5}$$

To map the cardinal demagnetization field components to the NV axis (Figure S4.1), with polar angle $\theta_{NV}$ and azimuthal angle $\phi_{NV}$, components $B_x$ and $B_z$ are projected to the NV orientation unit vector, $[\sin\theta_{NV}\cos\phi_{NV}, \sin\theta_{NV}\sin\phi_{NV}, \cos\theta_{NV}]$:

$$B_\parallel^{edge}(x, d_{NV}) = \sin\theta_{NV}\cos\phi_{NV} B_x^{edge}(x) + \sin\theta_{NV}\sin\phi_{NV} B_y^{edge}(x) + \cos\theta_{NV} B_z^{edge}(x)$$
$$= \frac{\mu_0 M_s t}{2\pi} \frac{1}{x^2 + d_{NV}^2} (d_{NV}\sin\theta_{NV}\cos\phi_{NV} - x\cos\theta_{NV}) \quad \text{Eq. S6}$$

Finally, to map the stray field for both edges of the nanowire, we consider the mirror case at the other edge of the wire, and the total stray field being a linear combination of the two, where the width of the wire is denoted by $w$:[78]

$$B_\parallel(x) = B_\parallel^{edge}(x) - B_\parallel^{edge}(x + w) \quad \text{Eq. S7}$$

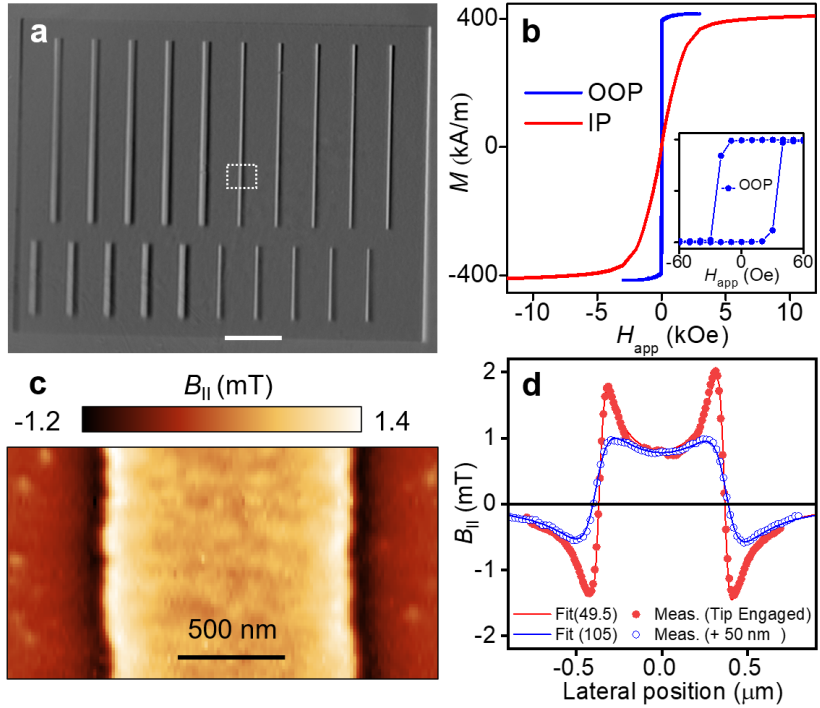

Figure S4.2. Characterization and device fabrication of the calibration sample. (a) SEM image of 4-nm thick $Co_3Pt$ patterned nanowires via EBL. The scale bar is 20 μm. (b) $M$–$H_{app}$ loops obtained by SQUID for IP and OOP magnetic field configurations. Inset of (b) shows the zoomed hysteresis loop for the OOP configuration, demonstrating PMA. (c) ODMR image of the stray field $B_\parallel$ at the edges of a perpendicularly magnetized $Co_3Pt$ wire. The scale bar is 500 nm. (d) Extracted $B_\parallel$ profiles at the edge of a wire for both AFM feedback engaged (red curve) and 50 nm additional stand-off (blue curve).

The calibration sample used was Pt (2 nm)/MgO (2 nm)/$Co_3Pt$ (4 nm)/STO, deposited by magnetron sputtering, with wires patterned by electron beam lithography (EBL) and Ar ion milling (Figure S4.2a) of dimensions 1.5 μm and 0.75 μm for width, and 50 μm for length. The $M$–$H_{app}$ hysteresis loop was measured by SQUID in both IP and OOP configurations (Figure S4.2b). We measured $M_S = 450$ kA/m and an effective magnetic anisotropy $K_{eff}$ of 1.54 mJ/m². For a wire of $w = 750$ nm, $t = 4$ nm, the ODMR image acquired with the tip feedback engaged is shown in Figure S4.2c. Several profiles were plotted for various controlled stand-off distances $d_{NV}$ as shown in Figure S4.2d. The line cut profiles (red and blue) of the measured stray field were then fit by equation S6 with corresponding lift heights of 49.5 nm and 105 nm, resulting in estimation of stand-off distance $d_{NV} = 49.5 \pm 3$ nm. Asymmetry in the fitted profile is a result of accounting for the difference in height of the NV sensor measuring over the wire at each edge, whose thickness inherently influences the stand-off, as well as the contribution from the slight deviation of $\phi_{NV}$ from 90°.

When using the same scanning probe for many measurements, the tip will inevitably accumulate some debris corresponding to an increase in $d_{NV}$. This effect can be monitored in the PL channel approach curve in Figure S4.3 and can be used as a reference for the NV stand-off from scan to scan. Oscillatory PL signal during the approach to a metallic surface has been reported in literature and is explained by the electromagnetic density of states near an interface,[38] which modulates the NV radiative lifetime. When the NV reaches < 100 nm from the metallic surface, stronger quenching dynamics take place which can reduce the PL to a fraction of its typical strength. The periodicity of this signal is consistent for a given interface and can thus be used to better estimate relative shifts in NV standoff $d_{NV}$. Of course, such estimations are only relative to the value that is obtained from an initial calibration procedure as described above and cannot be used as a conclusive stand-off measurement alone.

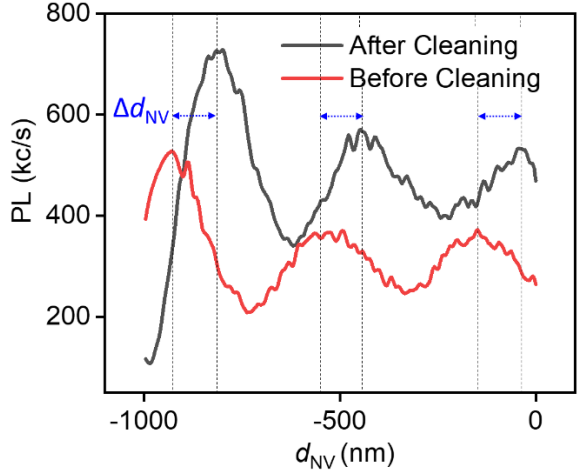

**Figure S4.3.** PL quenching approach curve before cleaning (red) and after cleaning (black) using patterned Si needles sample (TGT1, NT-MDT). Dotted lines show the relative distance ($d_{NV}$ = 120 nm) to the surface before AFM feedback engagement.

## S5. Vector reconstruction of skyrmion stray field components

For measurements of the stray-field $B_\parallel$ produced by skyrmions in the $(x, y)$ measurement plane at a distance $d_{NV}$ above an isolated skyrmion, $\mathbf{B}$ can be described by $-\nabla \phi_M$, where $\phi_M$ is the magnetostatic potential, and fields contained in this plane are source free. In Fourier space, the wave vector components of stray fields are linearly dependent with relationships governed by Gauss's law and Ampere's circuit law and described by taking the 2D Fourier transform, $\mathcal{F}(\mathbf{B}(B_x, B_y, B_z)) = \mathbf{b}(b_x, b_y, b_z)$, with:[80]

$$(ik_y)b_z(k_x, k_y, z) - \frac{\partial b_y(k_x, k_y, z)}{\partial z} = 0 \quad \text{Eq. S8}$$

$$(ik_x)b_z(k_x, k_y, z) - \frac{\partial b_x(k_x, k_y, z)}{\partial z} = 0 \quad \text{Eq. S9}$$

$$(ik_y)b_x(k_x, k_y, z) - (ik_x)b_y(k_x, k_y, z) = 0 \quad \text{Eq. S10}$$

Following analytical procedures outlined in reference [80], the following simplified relationships between the Fourier pair components of $\mathbf{b}$ are obtained:

$$b_y(k_x, k_y, z) = -(i\frac{k_y}{k})b_z(k_x, k_y, z) \quad \text{Eq. S11}$$

$$b_x(k_x, k_y, z) = -(i\frac{k_x}{k})b_z(k_x, k_y, z), \quad \text{Eq. S12}$$

where $k = \sqrt{k_y^2 + k_x^2}$, is the magnitude of the frequency vector. To introduce the effect of $d_{NV}$ in NV magnetometry measurements, the Hann window function is applied as a convolution in Fourier space, defined as follows:[40,49,52]

$$\text{hann}(k, d_{NV}) \begin{cases} \frac{1}{2}\left(1 + \cos\left(\frac{d_{NV}k}{2}\right)\right) & \text{if } \frac{d_{NV}k}{2p} > 1 \\ 0 & \text{if } \frac{d_{NV}k}{2p} < 1 \\ \text{Indeterminate} & \text{Otherwise} \end{cases} \quad \text{Eq. S13}$$

The Hann window filters out components of the image that have a frequency which is less than that defined by the resolution of the microscope, depending on the distance between the NV and the sample surface. The final reconstructed image is then obtained by performing the inverse Fourier transform. Reconstruction of $B_x$, $B_y$, and $B_z$ from the measured $B_\parallel$ in Figure 5a for $\Delta x = +50\%$ g-CoPt single layer is shown in Figure S5, matching the micromagnetic simulated maps (see Figures 5e and 5f of the simulated Bloch skyrmion in the main text).

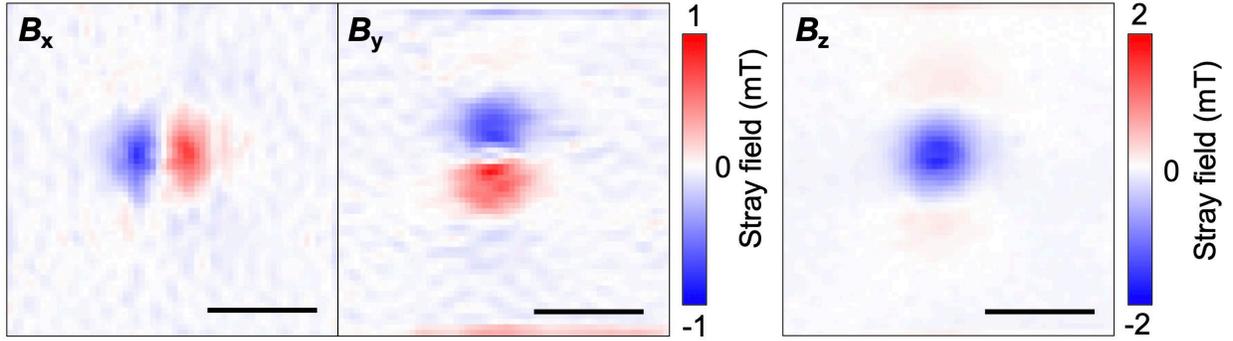

**Figure S5. Reconstruction of stray field components**. Reconstruction of magnetic stray field components from the measured $B_{//}$ in Figure 5a. The components are reconstructed at $d_{NV}$ = 320 nm above the surface. The scale bar is 500 nm.

## S6. Diamond probe height dependence in PL quenching imaging

In regimes of static magnetic field configurations which exceed several millitesla (mT), quantitative measurement of magnetic field is hindered due to the strong stray fields produced by the spin textures (skyrmions) in the transverse direction to the NV axis which cause spin mixing.[59] Perpendicular to the NV measurement axis, stray magnetic fields cause a steady decrease in the PL from the NV center. Off-axis magnetic fields also cause a reduction in ODMR contrast, preventing the acquisition of fully quantitative stray field images. Furthermore, level anti-crossings occurring in the excited and ground level occur when the field applied in the NV axis direction reaches ~51 mT and ~102 mT, which correlate to a narrow quenching distribution of the NV PL centered around the anti-crossing fields.[59]

For ferromagnetic samples with a relatively large saturation magnetization $M_s$, magnetic textures including domain walls and skyrmions can generate a strong static stray magnetic field (> 15 mT) near the sample surface which are too large to image by pixel-wise ODMR mapping. PL quenching can be used as an alternative imaging modality for such regimes, providing faster image acquisition at the cost of quantitative measurement of the field strength. In terms of modeling, the uncertainty of field orientation precludes only unidirectional reconstruction from simulated magnetic fields to PL quenching images.

The PL quenching image of a given magnetic texture will vary depending on the bias field applied while imaging, both due to the orientation of the bias field with respect to the NV sensing axis and the demagnetization field of interest. If the bias field is aligned with the core magnetization of a skyrmion, a significant portion of the demagnetization field will combine constructively with the bias field to induce further quenching. If the magnetic field is anti-parallel with the skyrmion core, the demagnetization field can oppose the bias field and can cancel out the quenching caused by the bias field. When the transverse fields reach twice the value of the bias field, further positive quenching dynamics begin to occur, as Zeeman splitting will have reached a negatively symmetric value. Within this framework, the configuration in which the bias field orientation opposes the magnetization of the skyrmion cores gives more detailed information about the relative strength of transverse fields present in the imaging plane. This case is shown in Figure S6.1, where PL quenching images are shown for a sequence of decreasing scanning heights. Initially, opposing magnetic fields to the bias field induce PL enhancement, which is broadened by the convolution introduced by the scanning NV distance. Drawing nearer to the skyrmion, field strength increases and eventually causes quenching in a pattern defined by the distribution of strong perpendicular stray field components.

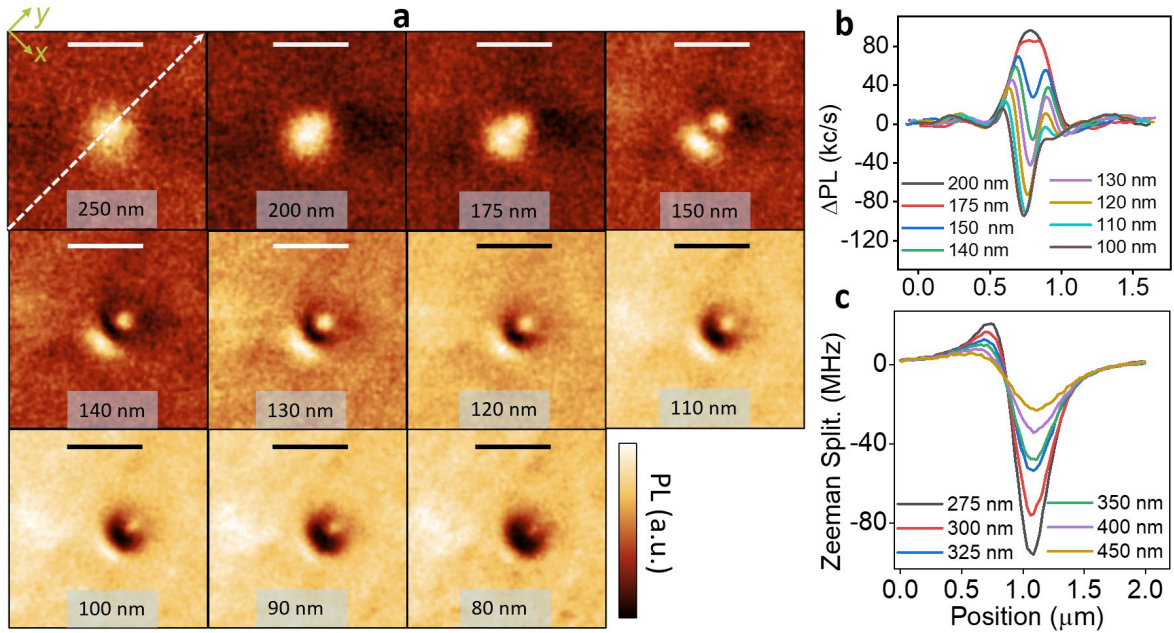

**Figure S6.1. Distance dependence of PL quenching imaging over isolated skyrmion in g-CoPt ($\Delta x$ = -50%) single layer**. (a) PL quenching image series in which $d_{NV}$ varies from 80 nm to 250 nm. The measured skyrmion is the same as shown in Figure 5a. The scale bar in all images in a is 500 nm. The scan frame was rotated clockwise by 45°. (b) Extracted $\Delta$PL line profiles are plotted together to show the evolution of quenching behavior with respect to $d_{NV}$. (c) Zeeman splitting profile measured at different $d_{NV}$ above an isolated skyrmion.

Because of the distance dependent resolution function associated with scanning probe magnetic imaging, it is common for the intricacies of magnetic features to be buried in the convolution with the distance dependent resolution function. In the case of pair skyrmions, scanning must take place at a height below a certain threshold above which their convolved signals appear as one gaussian blur. Figure S6.2 shows a similar series of PL quenching images as in Figure S6.1 for two skyrmion pairs. Even at heights of 150 nm, the stray fields generated by the skyrmions are strong enough to induce a significant PL reduction.

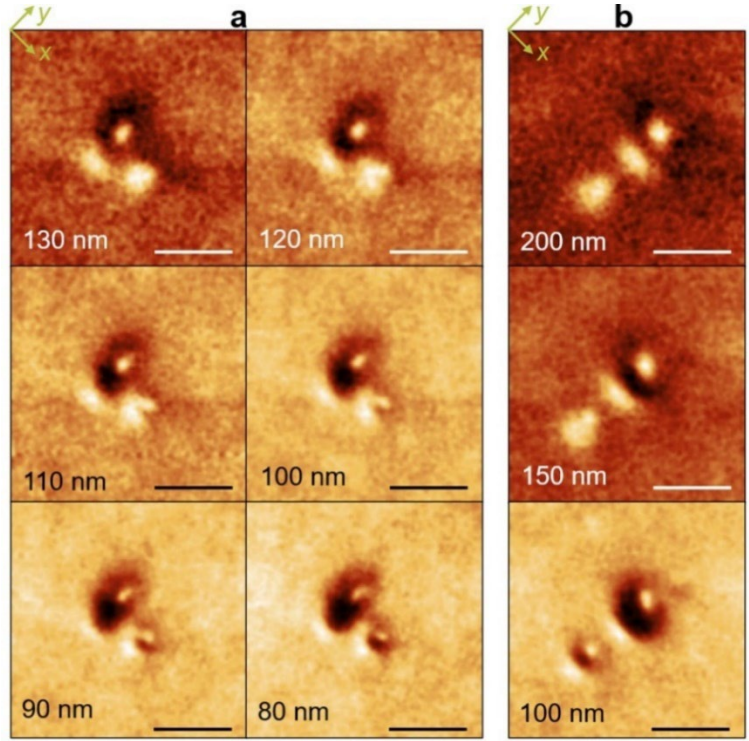

**Figure S6.2. PL quenching imaging over skyrmion pairs**. Representative PL image of differently oriented two skyrmion pairs (a) and (b) with respect to NV sensing axis. $d_{NV}$ is varied, specified by the distance in the lower left corner. The scale bar in (a-b) is 500 nm. The scan frame is rotated clockwise by 45°.

## S7. Corroboration of MFM and NV quenching imaging

Comparative analysis of NV and MFM imaging can be difficult due to the differences of the method of sensing magnetic fields and the relative constrictions imposed on measurement due to experimental limitations. In the case of MFM, the magnetic moment exerted on the sample by the tip can cause premature magnetization reversal if the tip magnetization is aligned antiparallel to the bulk sample magnetization orientation. In samples with PMA this can lead to a major discrepancy in the measured switched volume vs field when comparing with non-perturbative bulk magnetometry techniques. NV-SPM is non-perturbative and can be used to image hysteretic reversal without such influencing factors. However, the limitations of NV scanning magnetometry also hinder direct comparison. Sensor fragility and AFM modality have led to a standard stand-off $d_{NV}$ of ~ 50 nm, with further distance being preferred for safety and avoid degradation of signal due to strong quenching.

To confirm our imaging results, we compared a similar prepared magnetic state with corresponding MFM and NV quenching imaging, shown in Figure S7. The histograms of size distributions matched well implying that the back action of the MFM tip was not causing significant switching. The distribution in the NV quenching image displays larger features because the sensor is at an increased distance from the sample compared to MFM.

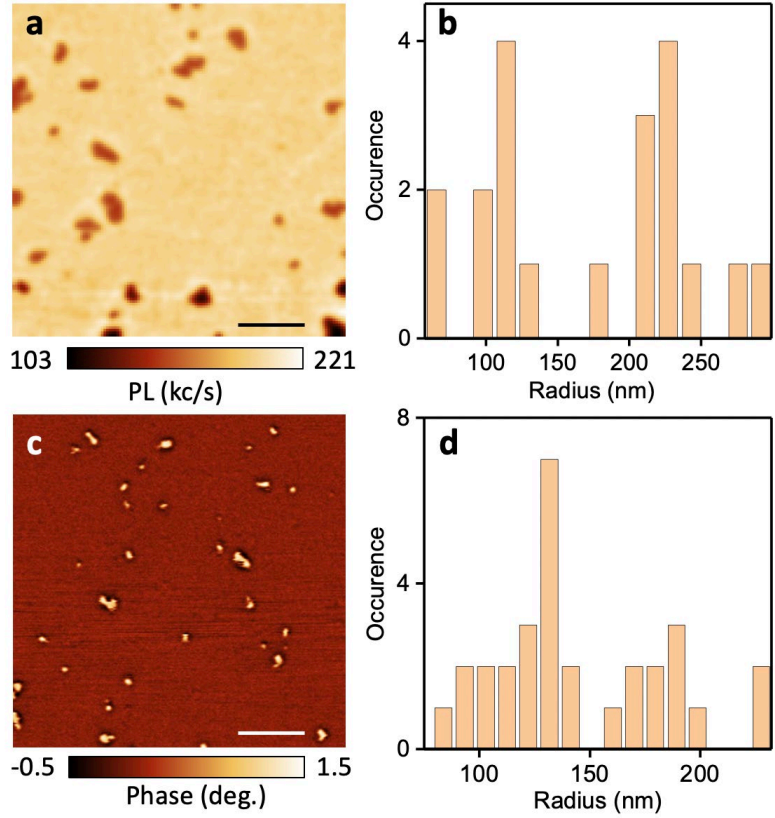

**Figure S7. Comparison of MFM and NV PL quenching imaging.** (a) NV PL quenching image of magnetic skyrmions. Skyrmions are nucleated from positive saturation (> 3 kOe) and imaged at -0.070 kOe. The image is obtained with the tip in AFM feedback ($d_{NV}$ ~ 50 nm). (b) Histogram of the measured radius of magnetic textures. MFM image (c) and histogram (d), obtained from a similarly representative state after similar field application protocols. Because this state corresponds to the transition to the spiral phase, there is a large distribution of radius, explained by the statistical nucleation of larger domain structures. The scale bar in (a) and (c) is 2 μm.

## S8. Micromagnetic simulations

To better understand the skyrmion pairs observed in the MFM and NV experiments we first performed a simulation at a temperature of 0 K by relaxing the system to equilibrium at zero magnetic field and gradually increased the applied magnetic field $H_{app}$.[34] Simulated images show the coexistence of Bloch skyrmions with other topological textures including antiskyrmion, higher order skyrmions and worm-like textures.[81,82] We start by relaxing the system to equilibrium at zero magnetic field (Figure S8.1a) and the spiral texture is stabilized with sparse isolated skyrmions. As the magnetic field $H_{app}$ is gradually increased to 0.3 kOe, as seen in Figure S8.1b, the spirals shrink into various types of spin textures. From the topological number density map in Figure S8.1c, near Bloch skyrmions (red circles) are the dominant texture, with antiskyrmions (blue circles), trivial bubbles (blue and red arcs), higher order skyrmions, and worm-like textures are also observed.

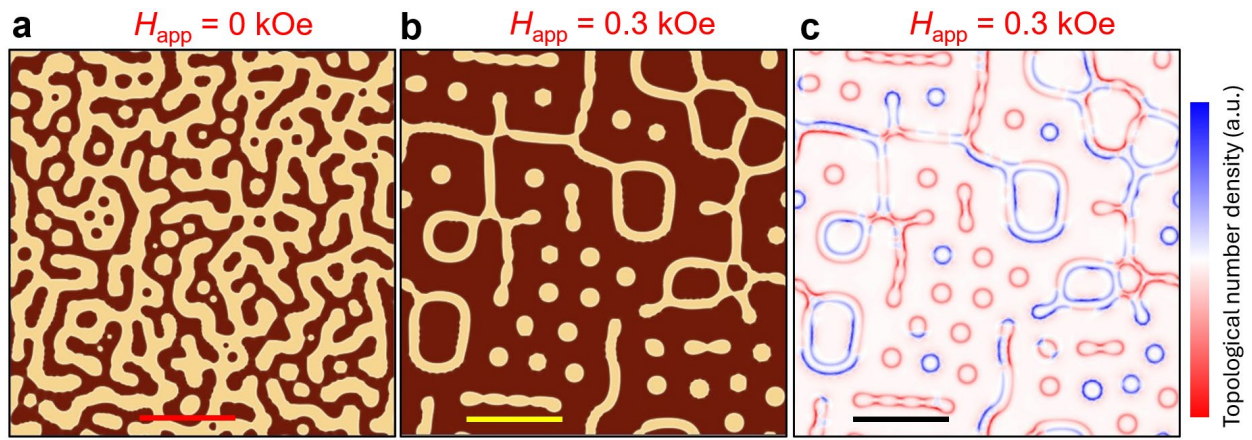

**Figure S8.1.** Micromagnetic simulation map of magnetic textures using negative gradient parameters at $H_{app}$ = 0 kOe (a) and 0.3 kOe (b). (c) Topological number density map of the image in (b) at $H_{app}$ = 0.3 kOe. The scale bar in a–c is 1 μm. Average DMI of 0.1 (mJ/m$^2$) is used at zero temperature. At zero magnetic field, the spiral textures are the dominating texture. When magnetic field (0.3 kOe) is applied, the spiral states shrink and stabilize into skyrmion (red circles in c), antiskyrmion (blue circles in c), trivial bubbles (blue and red arcs), and higher order skyrmions (winding number larger than one).

The domain wall angle of skyrmion is simulated by plotting its dependence with DMI and applied magnetic field $H_{app}$ for negative ($\Delta x$ = -50%, Figure S8.2a) and positive ($\Delta x$ = +50%, Figure S8.2b) g-CoPt 10 nm thick single layer at T = 0 K. By increasing DMI, the domain wall goes from pure Bloch ($\pi/2$) skyrmion to a hybrid domain wall angle. For our g-CoPt films with negative gradient DMI is ~ -0.5 mJ/m$^2$ and for positive gradient DMI is ~ 0.15 mJ/m$^2$, as measured from BLS spectroscopy [25] described in the main text, which matches the near-to-Bloch nature of the skyrmions imaged by NV and MFM.

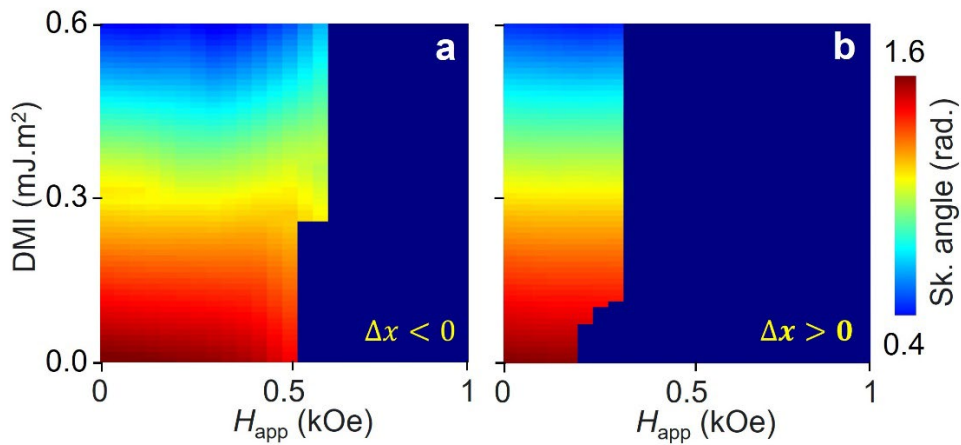

**Figure S8.2. Skyrmion helicity phase diagrams.** Calculated domain wall angle as a function of $H_{app}$ and average DMI for negative (a) and positive (b) gradient g-CoPt single layer (thickness = 10 nm). Domain wall angle of zero degrees corresponds to pure Néel skyrmion and $\pi/2$ corresponds to pure Bloch skyrmion. In the dark blue regions, the system is in a saturated FM state.

We further study the skyrmion-skyrmion and skyrmion-antiskyrmion pairs. Both types of pairs are possible states of the system. However, due to the large size of the skyrmion bubble cores, we see that both cases have repulsive force and require impurities to pin them. Finally, we simulated the system starting from a higher order skyrmion of winding number (W) of 2. We see that the system can also stabilize the higher order skyrmion. Figure S8.3 shows the micromagnetic simulations of skyrmion pairs and high order skyrmion (W = 2). In Figure S8.3a we show the granular structure of anisotropy used for pinning skyrmions and antiskyrmions in Figure S8.3b and Figure S8.3c, respectively. For both skyrmion-skyrmion (skm-skm) and skyrmion-antiskyrmion (skm-antiskm), the magnetic systems were first relaxed and then the granular structure was introduced. Higher order skyrmion is shown in Figure S8.3d. Next, we ran the Landau–Lifshitz–Gilbert (LLG) simulation for 200 ns at a temperature of 400 K to ensure they stay pinned in presence of thermal effects. The parameters used for the simulations here are for positive gradient g-CoPt films. Similar results were obtained with negative gradient g-CoPt film parameters as well. Average DMI of 0.1 mJ/m$^2$ is used. A Gilbert damping constant $\alpha$ of 0.1 is used for LLG part of the simulation.

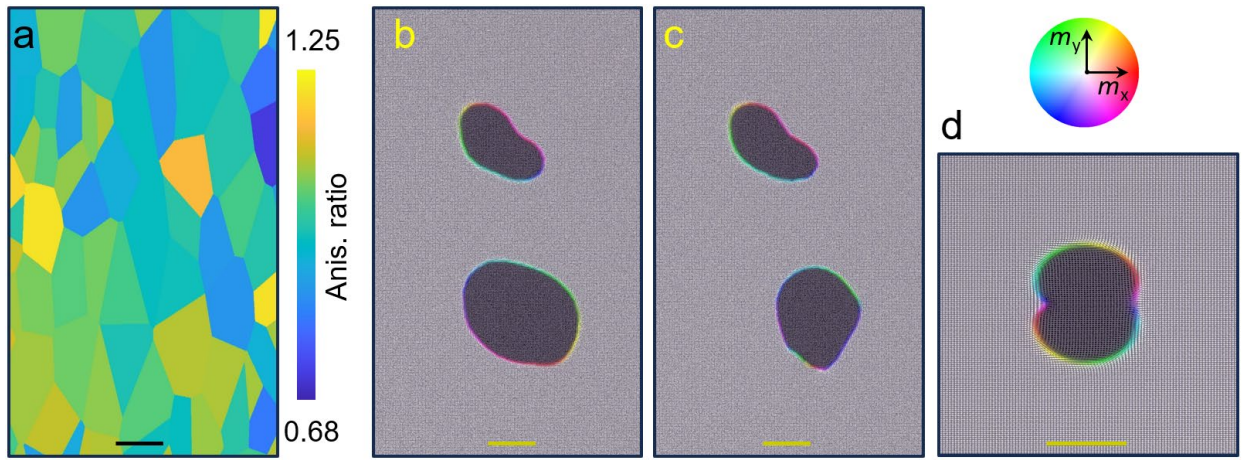

**Figure S8.3. Micromagnetic simulations of skyrmion-anti skyrmion, skyrmion-skyrmion pairs and higher order skyrmion.** (a) Underlying granular texture of g-CoP film used in simulated maps (b) and (c). The color bar shows the ratio of variations in anisotropy. Skyrmion-skyrmion pair (b) and skyrmion-antiskyrmion pair (c) relaxed in the granular texture. (d) Simulated skyrmion with winding number = 2. The scale bar in (a – d) is 200 nm. $H_{app}$ = 0.12 kOe in (b – c) and 0.18 kOe in (d).

Figure S8.4 shows the stray field projections of the simulated magnetization textures shown in Figure S8.3 at different measurement planes. Figures S8.4a and S8.4b show the stray field $B_\parallel$ projection of skm-antiskm pair and skm-skm pair at $d_{NV}$ = 150 nm, respectively, obtained by first relaxing the system of two skyrmions, then introducing the granular structure and running the LLG simulation for 200 ns at 400 K temperature. Due to the inclusion of pinning effects, the subsequent amorphousness would add an additional layer of complexity in attempting to distinguish the textures through reconstruction techniques. Moreover, in $d_{NV}$ regime where ODMR imaging on the skyrmion pairs is feasible (> 200 nm), the detailed shape is not able to be discerned due to convolution with the resolution function. Figure S8.4c shows the stray field $B_\parallel$ projections of skyrmion with W = 2 at $d_{NV}$ of 50 nm, 150 nm and 300 nm, respectively. At $d_{NV}$ = 150 nm and above, the $B_\parallel$ patterns closely resemble that of a single isolated skyrmion. For $d_{NV}$ > 150 nm, the stray fields seen here show similarities to the NV measured $B_\parallel$ images which could suggest that topological spin textures other than skyrmions can also exist in the sample (see main text). However, for $d_{NV} \leq 150$ nm the calculated stray-field produced by the skm-skm and skm-antiskm pairs, and W = 2 skyrmions is strong (≥ 15 mT) to perform NV ODMR imaging of such spin textures.

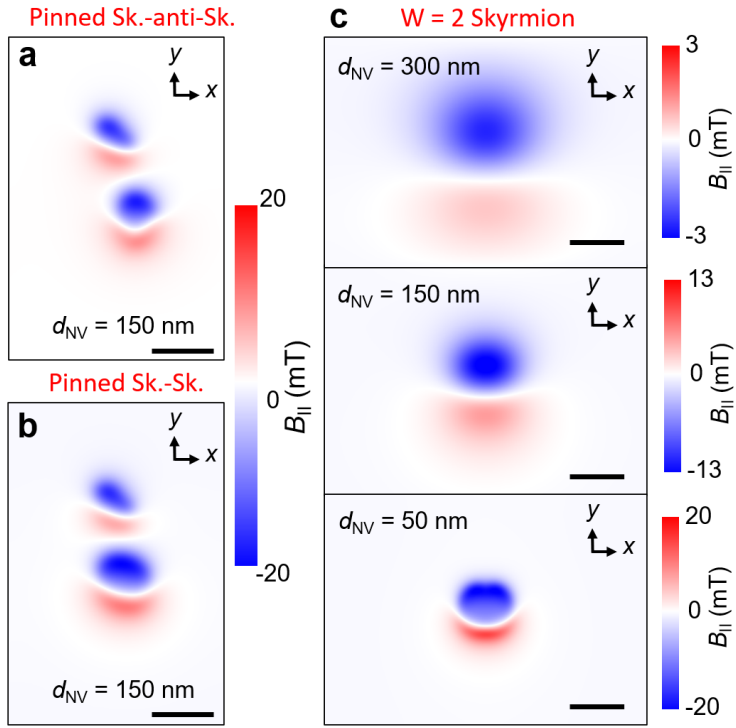

**Figure S8.4. Simulated NV ODMR images for pinned and higher order skyrmionic systems.** Calculated $B_\parallel$ maps for skm-antiskm pair (a) and skm-skm pair (b) at a $d_{NV}$ of 150 nm. The scale bar in (a-b) is 500 nm. (c) $B_\parallel$ maps for a higher order skyrmions (W = 2) calculated at $d_{NV}$ of 50, 150, and 300 nm, respectively. The scale bar in all images in (c) is 200 nm.

# REFERENCES

(1) Fert, A.; Reyren, N.; Cros, V. Magnetic Skyrmions: Advances in Physics and Potential Applications. *Nat Rev Mater* **2017**, *2* (7), 1–15. https://doi.org/10.1038/natrevmats.2017.31.
(2) Fert, A.; Cros, V.; Sampaio, J. Skyrmions on the Track. *Nature Nanotech* **2013**, *8*, 152–156. https://doi.org/10.1038/nnano.2013.29.
(3) Li, S.; Kang, W.; Huang, Y.; Zhang, X.; Zhou, Y.; Zhao, W. Magnetic Skyrmion-Based Artificial Neuron Device. *Nanotechnology* **2017**, *28*, 31LT01. https://doi.org/10.1088/1361-6528/aa7af5.
(4) Huang, Y.; Kang, W.; Zhang, X.; Zhou, Y.; Zhao, W. Magnetic Skyrmion-Based Synaptic Devices. *Nanotechnology* **2017**, *28*, 08LT02. https://doi.org/10.1088/1361-6528/aa5838.
(5) Moriya, T. Anisotropic Superexchange Interaction and Weak Ferromagnetism. *Phys. Rev.* **1960**, *120*, 91–98. https://doi.org/10.1103/PhysRev.120.91.
(6) Dzyaloshinsky, I. A Thermodynamic Theory of "Weak" Ferromagnetism of Antiferromagnetics. *Journal of Physics and Chemistry of Solids* **1958**, *4*, 241–255. https://doi.org/10.1016/0022-3697(58)90076-3.
(7) Soumyanarayanan, A.; Raju, M.; Gonzalez Oyarce, A. L.; Tan, A. K. C.; Im, M.-Y.; Petrović, A. P.; Ho, P.; Khoo, K. H.; Tran, M.; Gan, C. K.; Ernult, F.; Panagopoulos, C. Tunable Room-Temperature Magnetic Skyrmions in Ir/Fe/Co/Pt Multilayers. *Nature Mater* **2017**, *16*, 898–904. https://doi.org/10.1038/nmat4934.
(8) Heinze, S.; Von Bergmann, K.; Menzel, M.; Brede, J.; Kubetzka, A.; Wiesendanger, R.; Bihlmayer, G.; Blügel, S. Spontaneous Atomic-Scale Magnetic Skyrmion Lattice in Two Dimensions. *Nature Phys* **2011**, *7*, 713–718. https://doi.org/10.1038/nphys2045.

(9) Romming, N.; Hanneken, C.; Menzel, M.; Bickel, J. E.; Wolter, B.; Von Bergmann, K.; Kubetzka, A.; Wiesendanger, R. Writing and Deleting Single Magnetic Skyrmions. *Science* **2013**, *341*, 636–639. https://doi.org/10.1126/science.1240573.

(10) Moreau-Luchaire, C.; Moutafis, C.; Reyren, N.; Sampaio, J.; Vaz, C. A. F.; Van Horne, N.; Bouzehouane, K.; Garcia, K.; Deranlot, C.; Warnicke, P.; Wohlhüter, P.; George, J.-M.; Weigand, M.; Raabe, J.; Cros, V.; Fert, A. Additive Interfacial Chiral Interaction in Multilayers for Stabilization of Small Individual Skyrmions at Room Temperature. *Nature Nanotech* **2016**, *11*, 444–448. https://doi.org/10.1038/nnano.2015.313.

(11) Woo, S.; Litzius, K.; Krüger, B.; Im, M.-Y.; Caretta, L.; Richter, K.; Mann, M.; Krone, A.; Reeve, R. M.; Weigand, M.; Agrawal, P.; Lemesh, I.; Mawass, M.-A.; Fischer, P.; Kläui, M.; Beach, G. S. D. Observation of Room-Temperature Magnetic Skyrmions and Their Current-Driven Dynamics in Ultrathin Metallic Ferromagnets. *Nature Mater* **2016**, *15*, 501–506. https://doi.org/10.1038/nmat4593.

(12) Boulle, O.; Vogel, J.; Yang, H.; Pizzini, S.; De Souza Chaves, D.; Locatelli, A.; Menteş, T. O.; Sala, A.; Buda-Prejbeanu, L. D.; Klein, O.; Belmeguenai, M.; Roussigné, Y.; Stashkevich, A.; Chérif, S. M.; Aballe, L.; Foerster, M.; Chshiev, M.; Auffret, S.; Miron, I. M.; Gaudin, G. Room-Temperature Chiral Magnetic Skyrmions in Ultrathin Magnetic Nanostructures. *Nature Nanotech* **2016**, *11*, 449–454. https://doi.org/10.1038/nnano.2015.315.

(13) Li, Y.; Kanazawa, N.; Yu, X. Z.; Tsukazaki, A.; Kawasaki, M.; Ichikawa, M.; Jin, X. F.; Kagawa, F.; Tokura, Y. Robust Formation of Skyrmions and Topological Hall Effect Anomaly in Epitaxial Thin Films of MnSi. *Phys. Rev. Lett.* **2013**, *110*, 117202. https://doi.org/10.1103/PhysRevLett.110.117202.

(14) Yu, X. Z.; Onose, Y.; Kanazawa, N.; Park, J. H.; Han, J. H.; Matsui, Y.; Nagaosa, N.; Tokura, Y. Real-Space Observation of a Two-Dimensional Skyrmion Crystal. *Nature* **2010**, *465*, 901–904. https://doi.org/10.1038/nature09124.

(15) Yu, X. Z.; Kanazawa, N.; Onose, Y.; Kimoto, K.; Zhang, W. Z.; Ishiwata, S.; Matsui, Y.; Tokura, Y. Near Room-Temperature Formation of a Skyrmion Crystal in Thin-Films of the Helimagnet FeGe. *Nature Mater* **2011**, *10*, 106–109. https://doi.org/10.1038/nmat2916.

(16) Zheng, F.; Kiselev, N. S.; Yang, L.; Kuchkin, V. M.; Rybakov, F. N.; Blügel, S.; Dunin-Borkowski, R. E. Skyrmion–Antiskyrmion Pair Creation and Annihilation in a Cubic Chiral Magnet. *Nat. Phys.* **2022**, *18*, 863–868. https://doi.org/10.1038/s41567-022-01638-4.

(17) Wang, W.; Song, D.; Wei, W.; Nan, P.; Zhang, S.; Ge, B.; Tian, M.; Zang, J.; Du, H. Electrical Manipulation of Skyrmions in a Chiral Magnet. *Nat Commun* **2022**, *13*, 1593. https://doi.org/10.1038/s41467-022-29217-4.

(18) Zhang, Y.; Tang, J.; Wu, Y.; Shi, M.; Xu, X.; Wang, S.; Tian, M.; Du, H. Stable Skyrmion Bundles at Room Temperature and Zero Magnetic Field in a Chiral Magnet. *Nat Commun* **2024**, *15*, 3391. https://doi.org/10.1038/s41467-024-47730-6.

(19) Bogdanov, A. N.; Rößler, U. K.; Wolf, M.; Müller, K.-H. Magnetic Structures and Reorientation Transitions in Noncentrosymmetric Uniaxial Antiferromagnets. *Phys. Rev. B* **2002**, *66*, 214410. https://doi.org/10.1103/PhysRevB.66.214410.

(20) Vir, P.; Kumar, N.; Borrmann, H.; Jamijansuren, B.; Kreiner, G.; Shekhar, C.; Felser, C. Tetragonal Superstructure of the Antiskyrmion Hosting Heusler Compound $Mn_{1.4}PtSn$. *Chem. Mater.* **2019**, *31*, 5876–5880. https://doi.org/10.1021/acs.chemmater.9b02013.

(21) Ma, T.; Sharma, A. K.; Saha, R.; Srivastava, A. K.; Werner, P.; Vir, P.; Kumar, V.; Felser, C.; Parkin, S. S. P. Tunable Magnetic Antiskyrmion Size and Helical Period from Nanometers to Micrometers in a $D_{2d}$ Heusler Compound. *Advanced Materials* **2020**, *32*, 2002043. https://doi.org/10.1002/adma.202002043.

(22) Peng, L.; Takagi, R.; Koshibae, W.; Shibata, K.; Nakajima, K.; Arima, T.; Nagaosa, N.; Seki, S.; Yu, X.; Tokura, Y. Controlled Transformation of Skyrmions and Antiskyrmions in a Non-Centrosymmetric Magnet. *Nat. Nanotechnol.* **2020**, *15*, 181–186. https://doi.org/10.1038/s41565-019-0616-6.

(23) Nayak, A. K.; Kumar, V.; Ma, T.; Werner, P.; Pippel, E.; Sahoo, R.; Damay, F.; Rößler, U. K.; Felser, C.; Parkin, S. S. P. Magnetic Antiskyrmions above Room Temperature in Tetragonal Heusler Materials. *Nature* **2017**, *548*, 561–566. https://doi.org/10.1038/nature23466.

(24) Liang, J.; Chshiev, M.; Fert, A.; Yang, H. Gradient-Induced Dzyaloshinskii–Moriya Interaction. *Nano Lett.* **2022**, *22*, 10128–10133. https://doi.org/10.1021/acs.nanolett.2c03973.

(25) Zhang, Q.; Liang, J.; Bi, K.; Zhao, L.; Bai, H.; Cui, Q.; Zhou, H.-A.; Bai, H.; Feng, H.; Song, W.; Chai, G.; Gladii, O.; Schultheiss, H.; Zhu, T.; Zhang, J.; Peng, Y.; Yang, H.; Jiang, W. Quantifying the Dzyaloshinskii-Moriya Interaction Induced by the Bulk Magnetic Asymmetry. *Phys. Rev. Lett.* **2022**, *128*, 167202. https://doi.org/10.1103/PhysRevLett.128.167202.

(26) Zheng, Z.; Zhang, Y.; Lopez-Dominguez, V.; Sánchez-Tejerina, L.; Shi, J.; Feng, X.; Chen, L.; Wang, Z.; Zhang, Z.; Zhang, K.; Hong, B.; Xu, Y.; Zhang, Y.; Carpentieri, M.; Fert, A.; Finocchio, G.; Zhao, W.; Khalili Amiri, P. Field-Free Spin-Orbit Torque-Induced Switching of Perpendicular Magnetization in a Ferrimagnetic Layer with a Vertical Composition Gradient. *Nat Commun* **2021**, *12*, 4555. https://doi.org/10.1038/s41467-021-24854-7.

(27) Liu, L.; Zhou, C.; Zhao, T.; Yao, B.; Zhou, J.; Shu, X.; Chen, S.; Shi, S.; Xi, S.; Lan, D.; Lin, W.; Xie, Q.; Ren, L.; Luo, Z.; Sun, C.; Yang, P.; Guo, E.-J.; Dong, Z.; Manchon, A.; Chen, J. Current-Induced Self-Switching of Perpendicular Magnetization in CoPt Single Layer. *Nat Commun* **2022**, *13*, 3539. https://doi.org/10.1038/s41467-022-31167-w.

(28) Fert, A.; Levy, P. M. Role of Anisotropic Exchange Interactions in Determining the Properties of Spin-Glasses. *Phys. Rev. Lett.* **1980**, *44*, 1538–1541. https://doi.org/10.1103/PhysRevLett.44.1538.

(29) Rajput, N. S.; Sloyan, K.; Anjum, D. H.; Chiesa, M.; Ghaferi, A. A. A User-Friendly FIB Lift-out Technique to Prepare Plan-View TEM Sample of 2D Thin Film Materials. *Ultramicroscopy* **2022**, *235*, 113496. https://doi.org/10.1016/j.ultramic.2022.113496.

(30) Cortés-Ortuño, D.; Landeros, P. Influence of the Dzyaloshinskii–Moriya Interaction on the Spin-Wave Spectra of Thin Films. *J. Phys.: Condens. Matter* **2013**, *25*, 156001. https://doi.org/10.1088/0953-8984/25/15/156001.

(31) Belmeguenai, M.; Adam, J.-P.; Roussigné, Y.; Eimer, S.; Devolder, T.; Kim, J.-V.; Cherif, S. M.; Stashkevich, A.; Thiaville, A. Interfacial Dzyaloshinskii-Moriya Interaction in Perpendicularly Magnetized Pt/Co/AlO x Ultrathin Films Measured by Brillouin Light Spectroscopy. *Phys. Rev. B* **2015**, *91*, 180405. https://doi.org/10.1103/PhysRevB.91.180405.

(32) Di, K.; Zhang, V. L.; Lim, H. S.; Ng, S. C.; Kuok, M. H.; Qiu, X.; Yang, H. Asymmetric Spin-Wave Dispersion Due to Dzyaloshinskii-Moriya Interaction in an Ultrathin Pt/CoFeB Film. *Applied Physics Letters* **2015**, *106*, 052403. https://doi.org/10.1063/1.4907173.

(33) Zheng, F.; Li, H.; Wang, S.; Song, D.; Jin, C.; Wei, W.; Kovács, A.; Zang, J.; Tian, M.; Zhang, Y.; Du, H.; Dunin-Borkowski, R. E. Direct Imaging of a Zero-Field Target Skyrmion and Its Polarity Switch in a Chiral Magnetic Nanodisk. *Phys. Rev. Lett.* **2017**, *119*, 197205. https://doi.org/10.1103/PhysRevLett.119.197205.

(34) Raju, M.; Yagil, A.; Soumyanarayanan, A.; Tan, A. K. C.; Almoalem, A.; Ma, F.; Auslaender, O. M.; Panagopoulos, C. The Evolution of Skyrmions in Ir/Fe/Co/Pt Multilayers and Their Topological Hall Signature. *Nat Commun* **2019**, *10*, 696. https://doi.org/10.1038/s41467-018-08041-9.

(35) Liou, S.-H. Advanced Magnetic Force Microscopy Tips for Imaging Domains. In *Handbook of Advanced Magnetic Materials*; Liu, Y., Sellmyer, D. J., Shindo, D., Eds.; Springer US: Boston, MA, 2006; pp 374–396. https://doi.org/10.1007/1-4020-7984-2_10.

(36) Casiraghi, A.; Corte-León, H.; Vafaee, M.; Garcia-Sanchez, F.; Durin, G.; Pasquale, M.; Jakob, G.; Kläui, M.; Kazakova, O. Individual Skyrmion Manipulation by Local Magnetic Field Gradients. *Commun Phys* **2019**, *2*, 145. https://doi.org/10.1038/s42005-019-0242-5.

(37) Laraoui, A.; Hodges, J. S.; Meriles, C. A. Magnetometry of Random Ac Magnetic Fields Using a Single Nitrogen-Vacancy Center. *Appl. Phys. Lett.* **2010**, *97*, 143104. https://doi.org/10.1063/1.3497004.

(38) Maletinsky, P.; Hong, S.; Grinolds, M. S.; Hausmann, B.; Lukin, M. D.; Walsworth, R. L.; Loncar, M.; Yacoby, A. A Robust Scanning Diamond Sensor for Nanoscale Imaging with Single Nitrogen-Vacancy Centres. *Nature Nanotech* **2012**, *7*, 320–324. https://doi.org/10.1038/nnano.2012.50.

(39) Rondin, L.; Tetienne, J.-P.; Hingant, T.; Roch, J.-F.; Maletinsky, P.; Jacques, V. Magnetometry with Nitrogen-Vacancy Defects in Diamond. *Rep. Prog. Phys.* **2014**, *77*, 056503. https://doi.org/10.1088/0034-4885/77/5/056503.

(40) Casola, F.; Van Der Sar, T.; Yacoby, A. Probing Condensed Matter Physics with Magnetometry Based on Nitrogen-Vacancy Centres in Diamond. *Nat Rev Mater* **2018**, *3*, 17088. https://doi.org/10.1038/natrevmats.2017.88.

(41) Laraoui, A.; Ambal, K. Opportunities for Nitrogen-Vacancy-Assisted Magnetometry to Study Magnetism in 2D van Der Waals Magnets. *Applied Physics Letters* **2022**, *121*, 060502. https://doi.org/10.1063/5.0091931.

(42) Timalsina, R.; Wang, H.; Giri, B.; Erickson, A.; Xu, X.; Laraoui, A. Mapping of Spin-Wave Transport in Thulium Iron Garnet Thin Films Using Diamond Quantum Microscopy. *Adv Elect Materials* **2023**, 2300648. https://doi.org/10.1002/aelm.202300648.

(43) Lamichhane, S.; McElveen, K. A.; Erickson, A.; Fescenko, I.; Sun, S.; Timalsina, R.; Guo, Y.; Liou, S.-H.; Lai, R. Y.; Laraoui, A. Nitrogen-Vacancy Magnetometry of Individual Fe-Triazole Spin Crossover Nanorods. *ACS Nano* **2023**, *17*, 8694–8704. https://doi.org/10.1021/acsnano.3c01819.

(44) Doherty, M. W.; Manson, N. B.; Delaney, P.; Jelezko, F.; Wrachtrup, J.; Hollenberg, L. C. L. The Nitrogen-Vacancy Colour Centre in Diamond. *Physics Reports* **2013**, *528*, 1–45. https://doi.org/10.1016/j.physrep.2013.02.001.

(45) Thiel, L.; Rohner, D.; Ganzhorn, M.; Appel, P.; Neu, E.; Müller, B.; Kleiner, R.; Koelle, D.; Maletinsky, P. Quantitative Nanoscale Vortex Imaging Using a Cryogenic Quantum Magnetometer. *Nature Nanotech* **2016**, *11*, 677–681. https://doi.org/10.1038/nnano.2016.63.

(46) Appel, P.; Shields, B. J.; Kosub, T.; Hedrich, N.; Hübner, R.; Faßbender, J.; Makarov, D.; Maletinsky, P. Nanomagnetism of Magnetoelectric Granular Thin-Film Antiferromagnets. *Nano Lett.* **2019**, *19* (3), 1682–1687. https://doi.org/10.1021/acs.nanolett.8b04681.

(47) Thiel, L.; Wang, Z.; Tschudin, M. A.; Rohner, D.; Gutiérrez-Lezama, I.; Ubrig, N.; Gibertini, M.; Giannini, E.; Morpurgo, A. F.; Maletinsky, P. Probing Magnetism in 2D Materials at the Nanoscale with Single-Spin Microscopy. *Science* **2019**, *364 (6444)*, 973– 976,. https://doi.org/10.1126/science.aav6926.

(48) Finco, A.; Haykal, A.; Tanos, R.; Fabre, F.; Chouaieb, S.; Akhtar, W.; Robert-Philip, I.; Legrand, W.; Ajejas, F.; Bouzehouane, K.; Reyren, N.; Devolder, T.; Adam, J.-P.; Kim, J.-V.; Cros, V.; Jacques, V. Imaging Non-Collinear Antiferromagnetic Textures via Single Spin Relaxometry. *Nat Commun* **2021**, *12*, 767. https://doi.org/10.1038/s41467-021-20995-x.

(49) Erickson, A.; Shah, S. Q. A.; Mahmood, A.; Fescenko, I.; Timalsina, R.; Binek, C.; Laraoui, A. Nanoscale Imaging of Antiferromagnetic Domains in Epitaxial Films of $Cr_2O_3$ via Scanning Diamond Magnetic Probe Microscopy. *RSC Adv.* **2022**, *13*, 178–185. https://doi.org/10.1039/D2RA06440E.

(50) Zhou, T. X.; Carmiggelt, J. J.; Gächter, L. M.; Esterlis, I.; Sels, D.; Stöhr, R. J.; Du, C.; Fernandez, D.; Rodriguez-Nieva, J. F.; Büttner, F.; Demler, E.; Yacoby, A. A Magnon Scattering Platform. *Proceedings of the National Academy of Sciences* **2021**, *118*, e2019473118. https://doi.org/10.1073/pnas.2019473118.

(51) Simon, B. G.; Kurdi, S.; Carmiggelt, J. J.; Borst, M.; Katan, A. J.; van der Sar, T. Filtering and Imaging of Frequency-Degenerate Spin Waves Using Nanopositioning of a Single-Spin Sensor. *Nano Lett.* **2022**, *22*, 9198–9204. https://doi.org/10.1021/acs.nanolett.2c02791.

(52) Dovzhenko, Y.; Casola, F.; Schlotter, S.; Zhou, T. X.; Büttner, F.; Walsworth, R. L.; Beach, G. S. D.; Yacoby, A. Magnetostatic Twists in Room-Temperature Skyrmions Explored by Nitrogen-Vacancy Center Spin Texture Reconstruction. *Nat Commun* **2018**, *9*, 2712. https://doi.org/10.1038/s41467-018-05158-9.

(53) Rana, K. G.; Finco, A.; Fabre, F.; Chouaieb, S.; Haykal, A.; Buda-Prejbeanu, L. D.; Fruchart, O.; Le Denmat, S.; David, P.; Belmeguenai, M.; Denneulin, T.; Dunin-Borkowski, R. E.; Gaudin, G.;

Jacques, V.; Boulle, O. Room-Temperature Skyrmions at Zero Field in Exchange-Biased Ultrathin Films. *Phys. Rev. Applied* **2020**, *13*, 044079. https://doi.org/10.1103/PhysRevApplied.13.044079.

(54) Gross, I.; Martínez, L. J.; Tetienne, J.-P.; Hingant, T.; Roch, J.-F.; Garcia, K.; Soucaille, R.; Adam, J. P.; Kim, J.-V.; Rohart, S.; Thiaville, A.; Torrejon, J.; Hayashi, M.; Jacques, V. Direct Measurement of Interfacial Dzyaloshinskii-Moriya Interaction in X | CoFeB | MgO Heterostructures with a Scanning NV Magnetometer (X = Ta , TaN , and W). *Phys. Rev. B* **2016**, *94*, 064413. https://doi.org/10.1103/PhysRevB.94.064413.

(55) Akhtar, W.; Hrabec, A.; Chouaieb, S.; Haykal, A.; Gross, I.; Belmeguenai, M.; Gabor, M. S.; Shields, B.; Maletinsky, P.; Thiaville, A.; Rohart, S.; Jacques, V. Current-Induced Nucleation and Dynamics of Skyrmions in a Co -Based Heusler Alloy. *Phys. Rev. Applied* **2019**, *11*, 034066. https://doi.org/10.1103/PhysRevApplied.11.034066.

(56) Laraoui, A.; Aycock-Rizzo, H.; Gao, Y.; Lu, X.; Riedo, E.; Meriles, C. A. Imaging Thermal Conductivity with Nanoscale Resolution Using a Scanning Spin Probe. *Nat Commun* **2015**, *6*, 8954. https://doi.org/10.1038/ncomms9954.

(57) Erickson, A.; Shah, S. Q. A.; Mahmood, A.; Buragohain, P.; Fescenko, I.; Gruverman, A.; Binek, C.; Laraoui, A. Imaging Local Effects of Voltage and Boron Doping on Spin Reversal in Antiferromagnetic Magnetoelectric $Cr_2O_3$ Thin Films and Devices. *Adv Funct Materials* **2024**, 2408542. https://doi.org/10.1002/adfm.202408542.

(58) Appel, P.; Neu, E.; Ganzhorn, M.; Barfuss, A.; Batzer, M.; Gratz, M.; Tschöpe, A.; Maletinsky, P. Fabrication of All Diamond Scanning Probes for Nanoscale Magnetometry. *Review of Scientific Instruments* **2016**, *87*, 063703. https://doi.org/10.1063/1.4952953.

(59) Tetienne, J.-P.; Rondin, L.; Spinicelli, P.; Chipaux, M.; Debuisschert, T.; Roch, J.-F.; Jacques, V. Magnetic-Field-Dependent Photodynamics of Single NV Defects in Diamond: An Application to Qualitative All-Optical Magnetic Imaging. *New J. Phys.* **2012**, *14*, 103033. https://doi.org/10.1088/1367-2630/14/10/103033.

(60) Kovalev, A. A.; Sandhoefner, S. Skyrmions and Antiskyrmions in Quasi-Two-Dimensional Magnets. *Front. Phys.* **2018**, *6*, 98. https://doi.org/10.3389/fphy.2018.00098.

(61) Güngördü, U.; Nepal, R.; Tretiakov, O. A.; Belashchenko, K.; Kovalev, A. A. Stability of Skyrmion Lattices and Symmetries of Quasi-Two-Dimensional Chiral Magnets. *Phys. Rev. B* **2016**, *93*, 064428. https://doi.org/10.1103/PhysRevB.93.064428.

(62) Koshibae, W.; Nagaosa, N. Theory of Antiskyrmions in Magnets. *Nat Commun* **2016**, *7*, 10542. https://doi.org/10.1038/ncomms10542.

(63) Goerzen, M. A.; Von Malottki, S.; Meyer, S.; Bessarab, P. F.; Heinze, S. Lifetime of Coexisting Sub-10 Nm Zero-Field Skyrmions and Antiskyrmions. *npj Quantum Mater.* **2023**, *8*, 54. https://doi.org/10.1038/s41535-023-00586-3.

(64) Gruber, R.; Zázvorka, J.; Brems, M. A.; Rodrigues, D. R.; Dohi, T.; Kerber, N.; Seng, B.; Vafaee, M.; Everschor-Sitte, K.; Virnau, P.; Kläui, M. Skyrmion Pinning Energetics in Thin Film Systems. *Nat Commun* **2022**, *13*, 3144. https://doi.org/10.1038/s41467-022-30743-4.

(65) Göbel, B.; Henk, J.; Mertig, I. Forming Individual Magnetic Biskyrmions by Merging Two Skyrmions in a Centrosymmetric Nanodisk. *Sci Rep* **2019**, *9*, 9521. https://doi.org/10.1038/s41598-019-45965-8.

(66) Ajejas, F.; Sassi, Y.; Legrand, W.; Collin, S.; Peña Garcia, J.; Thiaville, A.; Pizzini, S.; Reyren, N.; Cros, V.; Fert, A. Interfacial Potential Gradient Modulates Dzyaloshinskii-Moriya Interaction in Pt/Co/Metal Multilayers. *Phys. Rev. Materials* **2022**, *6*, L071401. https://doi.org/10.1103/PhysRevMaterials.6.L071401.

(67) Chen, G. Skyrmion Hall Effect. *Nature Phys* **2017**, *13*, 112–113. https://doi.org/10.1038/nphys4030.

(68) Vélez, S.; Ruiz-Gómez, S.; Schaab, J.; Gradauskaite, E.; Wörnle, M. S.; Welter, P.; Jacot, B. J.; Degen, C. L.; Trassin, M.; Fiebig, M.; Gambardella, P. Current-Driven Dynamics and Ratchet Effect of Skyrmion Bubbles in a Ferrimagnetic Insulator. *Nat. Nanotechnol.* **2022**, *17*, 834–841. https://doi.org/10.1038/s41565-022-01144-x.

(69) Vansteenkiste, A.; Leliaert, J.; Dvornik, M.; Helsen, M.; Garcia-Sanchez, F.; Van Waeyenberge, B. The Design and Verification of MuMax3. *AIP Advances* **2014**, *4*, 107133. https://doi.org/10.1063/1.4899186.

(70) Chen, Z.; Liu, L.; Ye, Z.; Chen, Z.; Zheng, H.; Jia, W.; Zeng, Q.; Wang, N.; Xiang, B.; Lin, T.; Liu, J.; Qiu, M.; Li, S.; Shi, J.; Han, P.; An, H. Current-Induced Magnetization Switching in a Chemically Disordered A1 CoPt Single Layer. *Appl. Phys. Express* **2021**, *14*, 033002. https://doi.org/10.35848/1882-0786/abdcd5.

(71) Saito, S.; Nozawa, N.; Hinata, S.; Takahashi, M.; Shibuya, K.; Hoshino, K.; Awaya, S. Compositional Modulated Atomic Layer Stacking and Uniaxial Magnetocrystalline Anisotropy of CoPt Alloy Sputtered Films with Close-Packed Plane Orientation. *Journal of Applied Physics* **2015**, *117*, 17C753. https://doi.org/10.1063/1.4918760.

(72) Johnson, M. T.; Bloemen, P. J. H.; Broeder, F. J. A. D.; Vries, J. J. D. Magnetic Anisotropy in Metallic Multilayers. *Rep. Prog. Phys.* **1996**, *59*, 1409–1458. https://doi.org/10.1088/0034-4885/59/11/002.

(73) Kapusta, O.; Zelenakova, A.; Bednarcik, J.; Zelenak, V. Structural Transformation Study of CoPt Particles for High Density Recording Media; Slovak University of Technology: Slovakia, 2017; p 346.

(74) Ohtake, M.; Ouchi, S.; Kirino, F.; Futamoto, M. Structure and Magnetic Properties of CoPt, CoPd, FePt, and FePd Alloy Thin Films Formed on MgO(111) Substrates. *IEEE Transactions on Magnetics* **2012**, *48*, 3595–3598. https://doi.org/10.1109/TMAG.2012.2198875.

(75) Zeper, W. B.; Greidanus, F. J. A. M.; Carcia, P. F.; Fincher, C. R. Perpendicular Magnetic Anisotropy and Magneto-Optical Kerr Effect of Vapor-Deposited Co/Pt-Layered Structures. *Journal of Applied Physics* **1989**, *65*, 4971–4975. https://doi.org/10.1063/1.343189.

(76) Yang, Y.; Varghese, B.; Tan, H. K.; Wong, S. K.; Piramanayagam, S. N. Microstructure Investigations of Hcp Phase CoPt Thin Films with High Coercivity. *Journal of Applied Physics* **2014**, *115*, 083910. https://doi.org/10.1063/1.4866817.

(77) Balasubramanian, G.; Neumann, P.; Twitchen, D.; Markham, M.; Kolesov, R.; Mizuochi, N.; Isoya, J.; Achard, J.; Beck, J.; Tissler, J.; Jacques, V.; Hemmer, P. R.; Jelezko, F.; Wrachtrup, J. Ultralong Spin Coherence Time in Isotopically Engineered Diamond. *Nature Mater* **2009**, *8*, 383–387. https://doi.org/10.1038/nmat2420.

(78) Celano, U.; Zhong, H.; Ciubotaru, F.; Stoleriu, L.; Stark, A.; Rickhaus, P.; De Oliveira, F. F.; Munsch, M.; Favia, P.; Korytov, M.; Van Marcke, P.; Maletinsky, P.; Adelmann, C.; Van Der Heide, P. Probing Magnetic Defects in Ultra-Scaled Nanowires with Optically Detected Spin Resonance in Nitrogen-Vacancy Center in Diamond. *Nano Lett.* **2021**, *21*, 10409–10415. https://doi.org/10.1021/acs.nanolett.1c03723.

(79) Hingant, T.; Tetienne, J.-P.; Martínez, L. J.; Garcia, K.; Ravelosona, D.; Roch, J.-F.; Jacques, V. Measuring the Magnetic Moment Density in Patterned Ultrathin Ferromagnets with Submicrometer Resolution. *Phys. Rev. Applied* **2015**, *4*, 014003. https://doi.org/10.1103/PhysRevApplied.4.014003.

(80) Lima, E. A.; Weiss, B. P. Obtaining Vector Magnetic Field Maps from Single-Component Measurements of Geological Samples. *J. Geophys. Res.* **2009**, *114*, B06102. https://doi.org/10.1029/2008JB006006.

(81) Göbel, B.; Mertig, I.; Tretiakov, O. A. Beyond Skyrmions: Review and Perspectives of Alternative Magnetic Quasiparticles. *Physics Reports* **2021**, *895*, 1–28. https://doi.org/10.1016/j.physrep.2020.10.001.

(82) Yu, X. Z.; Tokunaga, Y.; Kaneko, Y.; Zhang, W. Z.; Kimoto, K.; Matsui, Y.; Taguchi, Y.; Tokura, Y. Biskyrmion States and Their Current-Driven Motion in a Layered Manganite. *Nat Commun* **2014**, *5*, 3198. https://doi.org/10.1038/ncomms4198.